\PassOptionsToPackage{table}{xcolor}
\documentclass[sigconf,natbib=true,anonymous=false]{acmart}

\usepackage{graphicx}
\usepackage{subcaption}
\usepackage{algorithm}
\usepackage{algpseudocode}
\usepackage{enumitem}
\usepackage{multirow}
\usepackage{tabularx}
\usepackage{threeparttable}
\usepackage{color, xspace}
\usepackage{appendix}
\usepackage[table]{xcolor}
\usepackage{tcolorbox}
\usepackage{booktabs}     
\usepackage{amsmath}      
\usepackage{array}        
\usepackage{caption}      

\definecolor{searchcolor}{RGB}{40, 120, 180}      
\definecolor{doccolor}{RGB}{160, 80, 40}          
\definecolor{factcolor}{RGB}{60, 140, 60}         
\definecolor{analcolor}{RGB}{140, 60, 140}        
\definecolor{anscolor}{RGB}{180, 20, 20}          

\newcommand{\tagsearch}[1]{\textcolor{searchcolor}{\texttt{<#1>}}}
\newcommand{\tagendsearch}[1]{\textcolor{searchcolor}{\texttt{</#1>}}}
\newcommand{\tagdoc}[1]{\textcolor{doccolor}{\texttt{<#1>}}}
\newcommand{\tagenddoc}[1]{\textcolor{doccolor}{\texttt{</#1>}}}
\newcommand{\tagfacts}[1]{\textcolor{factcolor}{\texttt{<#1>}}}
\newcommand{\tagendfacts}[1]{\textcolor{factcolor}{\texttt{</#1>}}}
\newcommand{\taganal}[1]{\textcolor{analcolor}{\texttt{<#1>}}}
\newcommand{\tagendanal}[1]{\textcolor{analcolor}{\texttt{</#1>}}}
\newcommand{\tagans}[1]{\textcolor{anscolor}{\texttt{<#1>}}}
\newcommand{\tagendans}[1]{\textcolor{anscolor}{\texttt{</#1>}}}

\AtBeginDocument{%
  }
\setcopyright{acmlicensed}
\copyrightyear{2018}
\acmYear{2018}
\acmDOI{XXXXXXX.XXXXXXX}
\acmConference[Conference acronym 'XX]{Make sure to enter the correct
  conference title from your rights confirmation email}{June 03--05,
  2018}{Woodstock, NY}
\acmISBN{978-1-4503-XXXX-X/2018/06}

\usepackage{color, xspace}

\begin{document}

\title{R$^2$-Searcher: Calibrating Retrieval and Reasoning Boundaries for Agentic Search}


\author{Sheng Zhang}
\affiliation{%
  \institution{City University of Hong Kong}
  \streetaddress{}
  \city{Hong Kong}
  \state{}
  \country{China}
}
\email{szhang844-c@my.cityu.edu.hk}

\author{Junyi Li}
\affiliation{%
  \institution{City University of Hong Kong}
  \streetaddress{}
  \city{Hong Kong}
  \state{}
  \country{China}
}
\email{junyili@cityu.edu.hk}

\author{Wenlin Zhang}
\affiliation{%
  \institution{City University of Hong Kong}
  \streetaddress{}
  \city{Hong Kong}
  \state{}
  \country{China}
}
\email{wl.z@my.cityu.edu.hk}

\author{Xiaowei Qian}
\affiliation{%
  \institution{City University of Hong Kong}
  \streetaddress{}
  \city{Hong Kong}
  \state{}
  \country{China}
}
\email{xiaowqian2-c@my.cityu.edu.hk}

\author{Yichao Wang}
\authornote{Corresponding Authors}
\affiliation{%
  \institution{Huawei Technologies Ltd.}
  \streetaddress{}
  \city{}
  \state{}
  \country{Signapore}
}
\email{wangyichao5@huawei.com}

\author{Yingyi Zhang}
\affiliation{%
  \institution{City University of Hong Kong and Dalian University of Technology}
  \streetaddress{}
  \city{}
  \state{}
  \country{Hong Kong, China}
}
\email{yingyizhang@mail.dlut.edu.cn}

\author{Maolin Wang}
\affiliation{%
  \institution{City University of Hong Kong}
  \streetaddress{}
  \city{}
  \state{}
  \country{Hong Kong, China}
}
\email{morin.wang@my.cityu.edu.hk}

\author{Yong Liu}
\affiliation{%
  \institution{Huawei Technologies Ltd.}
  \streetaddress{}
  \city{}
  \state{}
  \country{Signapore}
}
\email{liu.yong6@huawei.com}

\author{Xiangyu Zhao}
\authornotemark[1]
\affiliation{%
  \institution{City University of Hong Kong}
  \streetaddress{}
  \city{Hong Kong}
  \state{}
  \country{China}
  \postcode{}
}
\email{xianzhao@cityu.edu.hk}

\renewcommand{\shortauthors}{Sheng Zhang et al.}

\begin{abstract}

Recent search agents for multi-hop reasoning often fail by either retrieving incomplete evidence or reasoning over irrelevant portions of the retrieved content, leading to a retrieval-reasoning boundary shift. We propose R$^2$-Searcher, a novel framework that explicitly explores and calibrates the retrieval and reasoning boundaries via fine-grained, query-token-guided evidence modeling and post-retrieval reflection. Specifically, R$^2$-Searcher: (1) constructs fine-grained reasoning contexts by extracting precise facts from retrieved content based on query token semantics (e.g., subjects, actions, temporal markers, and degree modifiers), thereby guiding the attention of search agent; (2) introduces a retrieval reflection mechanism that evaluates and corrects boundary deviations after each retrieval step, guiding the generation of improved queries grounded in the extracted reasoning contexts; and (3) employs an end-to-end reasoning-reflection-guided reinforcement learning algorithm, R$^2$PO, which jointly optimizes both boundaries through a tree-based exploration of reasoning regions and reflections. Our method significantly enhances the quality of both retrieval and reasoning, establishing an iterative loop where retrieval and reasoning mutually enhance each other. Extensive experiments on seven complex multi-hop QA benchmarks demonstrate that R$^2$-Searcher significantly outperforms state-of-the-art agentic search methods in answer accuracy and retrieval-reasoning quality. 
Ablation studies further confirm the critical role of retrieval-reasoning boundary calibration.
Implementation code is available for reproducibility: https://github.com/szhang-cityu/R2-Searcher.
\end{abstract}




\maketitle

\section{Introduction}
In recent years, Retrieval-Augmented Generation (RAG) has emerged as an effective approach to leverage external knowledge and enhance the answer accuracy of Large Language Models (LLMs)~\cite{RAG}. The integration of such external knowledge has also become increasingly prevalent in the emerging field of LLM-based agents~\cite{lsrp, astute_rag, kgrag, hipporag2}. 
To overcome the limited retrieval depth of conventional RAG, recent works have developed search agents that extend both the depth of search and the complexity of reasoning~\cite{searcho1, searchr1}. 
Despite these advances, existing agentic search methods often fail to ensure that retrieved information consistently supports the evolving reasoning process, leading to brittle performance on complex multi-hop question answering (QA) tasks.


At a fundamental level, search agents suffer from a persistent \emph{misalignment} between retrieval and reasoning, where retrieved content does not reliably support the inference state required at each step of the search. We identify two concrete manifestations of this misalignment:
(1) \emph{Attention Dilution in Step-wise Observations}: LLM search agents receive lengthy and semantically complex documents in each search step, which causes critical evidence to be overlooked or misused~\cite{lost_middle, mem1}. This often leads to a shift in the reasoning boundary, where inference is no longer grounded in answer-relevant content.
(2) \emph{Extremely Large Query-Document Interaction Space}: 
The vast space of query–document transitions makes iterative retrieval highly unstable. Without explicit signals indicating whether retrieved evidence supports the current reasoning state, agents are prone to drifting toward irrelevant or misleading contexts~\cite{reasonrag}, compounding early errors over multiple steps~\cite{deepretrieval}.

\begin{figure}[t]
    \centering
    \includegraphics[width=\linewidth]{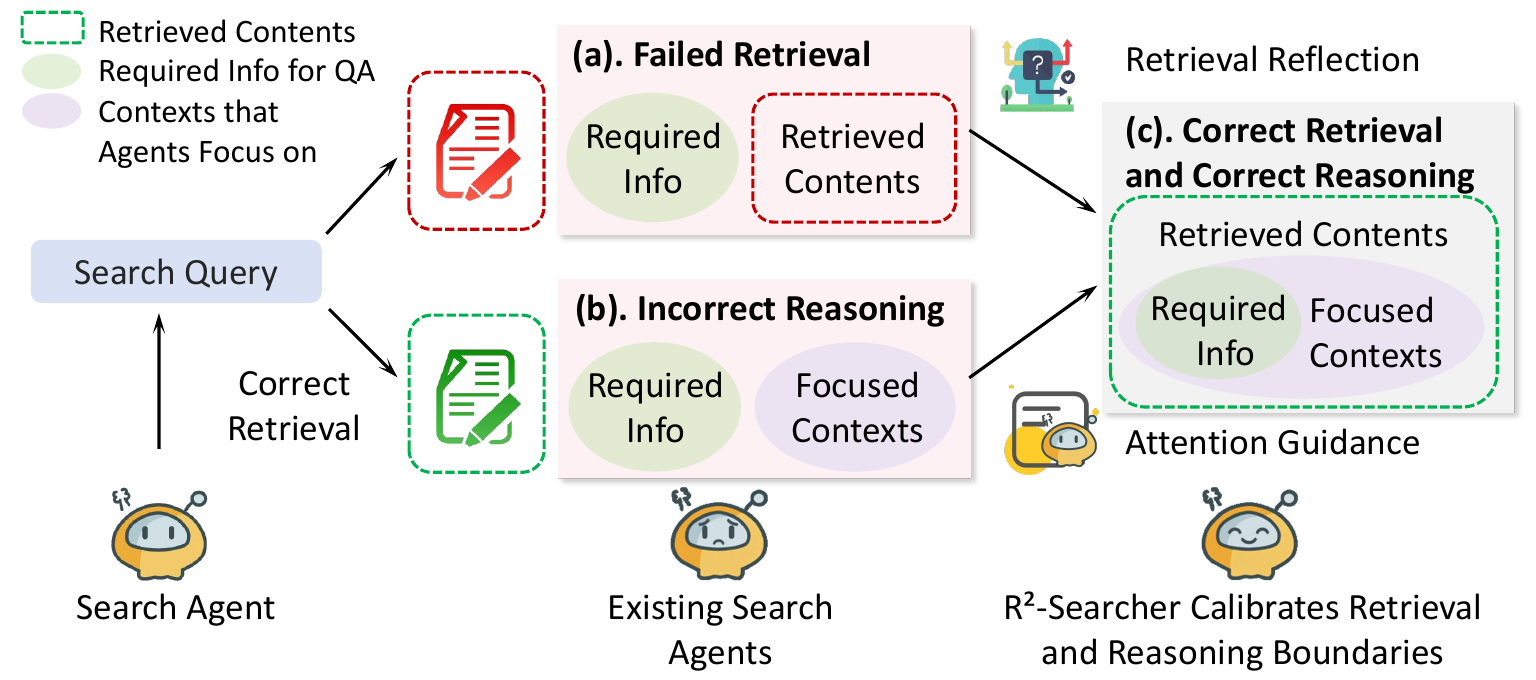}
    \vspace{-10pt}
    \caption{Existing search agents suffer from completely mismatched retrieval and incorrect reasoning, while R$^2$-Searcher calibrates the retrieval and reasoning boundaries.}
    \label{fig:challenges}
    \vspace{-5mm}
\end{figure}

Existing agentic search methods primarily attempt to mitigate these issues through two strategies: \emph{document summarization} and \emph{query enhancement}. However, both strategies operate in a loosely coupled manner with the reasoning process.
(1) Document summarization methods compress either individual documents~\cite{searcho1, amem, autorefine} or the accumulated retrieval history~\cite{mem1, supo, iterresearch, memsearcher} to help the agent focus on the information for answers. In this paper, we refer to contexts that the agent focuses on for reasoning as "reasoning regions". Yet, these summaries remain coarse-grained and rely on static heuristics that fail to guide the agent to the concise and correct reasoning region.
(2) Query enhancement methods aim to constrain the action space by reformulating queries prior to retrieval~\cite{deepretrieval}. Since such refinements are performed without access to the forthcoming observations, they are inherently incapable of identifying missing or misleading evidence. Although reinforcement learning–based query policies can improve retrieval quality~\cite{autorefine, searchr1}, they often suffer from hallucinated queries or reward-driven repetition.
Crucially, none of these approaches provides an \emph{explicit} mechanism to assess whether retrieved content supports the current reasoning requirement, leading to search trajectories that are trapped by incorrect reasoning or failed retrievals. We illustrate the comparison with previous work in Figure~\ref{fig:challenges}, showing the problem of reasoning-retrieval boundary shift in the agentic search process.

To address these challenges, we propose \textbf{R$^2$-Searcher}, an agentic search framework that explicitly calibrates the \textbf{R}easoning-\textbf{R}etrieval boundary for each search step, ensuring that retrieval is dynamically aligned with the agent's inference state. R$^2$-Searcher calibrates the reasoning region under the guidance of factual elements based on query token groups. 
Rather than retaining full documents or heuristic summaries, these factual elements help the agent focus on the region directly relevant to the current reasoning objective, grounding inference in concise and interpretable content.
To mitigate retrieval–reasoning mismatch, we enable the model to focus on and reflect on filtered contexts
, thereby guiding search toward answer-relevant directions.
Building on this formulation, we extend Group Relative Policy Optimization (GRPO)~\cite{grpo} and introduce Reasoning-Reflection Guided Policy Optimization (R$^2$PO), an end-to-end reinforcement learning algorithm that optimizes search,  reasoning region extraction, and reflection policies. To efficiently explore the coupled reasoning–retrieval action space and optimize the actions with process-level rewards, R$^2$PO employs tree-based rollouts that jointly consider candidate reasoning states and retrieval actions.
By reflecting on whether retrieved contents support the reasoning requirements to refine subsequent queries, R$^2$-Searcher establishes a closed-loop mechanism where retrieval quality and reasoning accuracy mutually reinforce each other. Our R$^2$-Searcher achieves explicit, dynamic co-optimization of the retrieval-reasoning boundary, enabling more accurate, faithful, and efficient agentic search.

\begin{figure*}[t]
    \centering
    \includegraphics[width=0.995\linewidth]{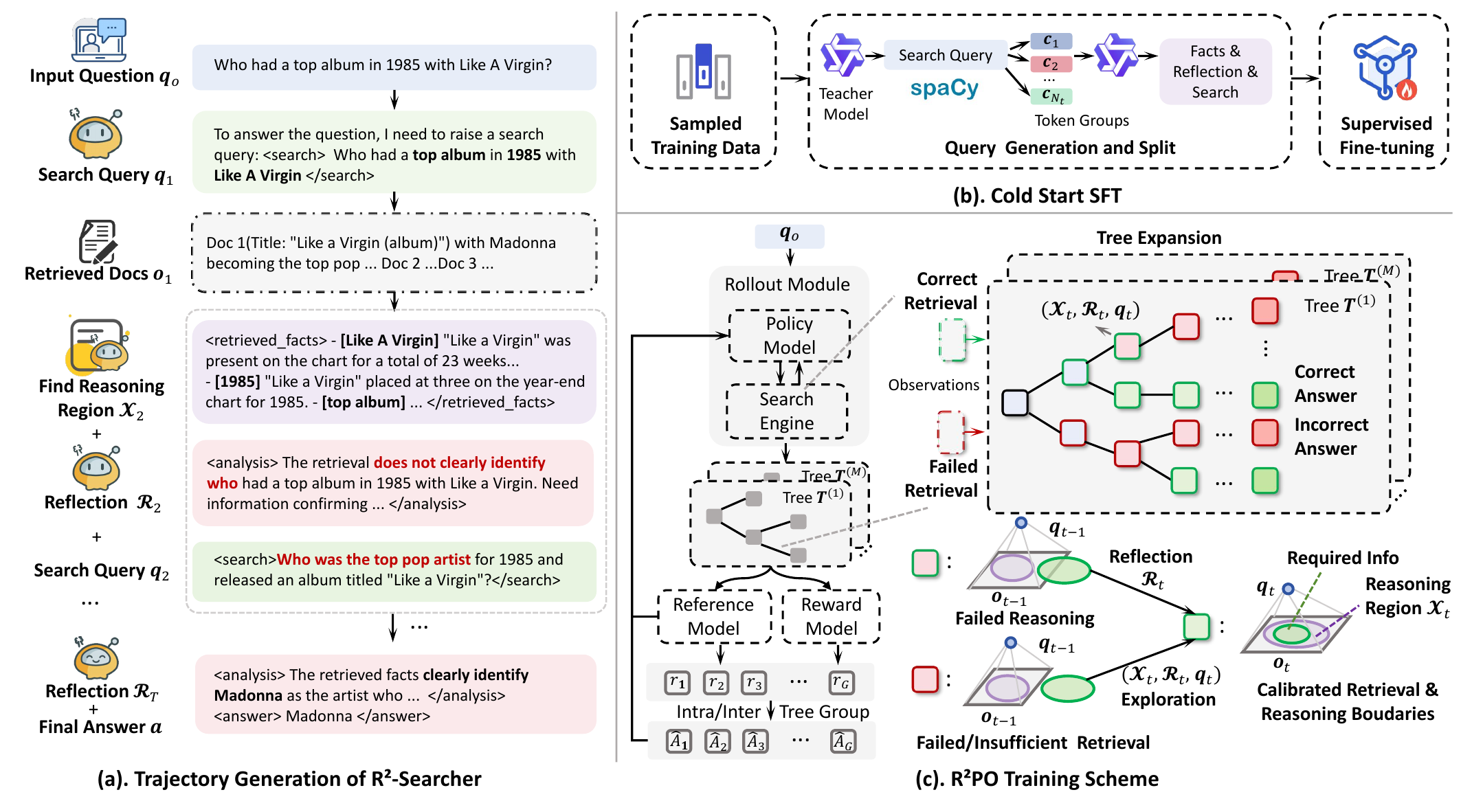}
    \vspace{-10pt}
    \caption{Overview of R$^2$-Searcher Trajectory Generation and Training Scheme. (a). The trajectory generation of R$^2$-Searcher with step-wise concise reasoning contexts and explicit reflection. (b) Data Preparation and SFT training for R$^2$-Searcher. (c) R$^2$PO training scheme where tree-based expansion is applied to explore retrieval-reasoning boundary calibration.}
    \label{fig:overview}
    \vspace{-1mm}
\end{figure*}
Our main contributions are summarized as follows:

\begin{itemize}[leftmargin=*]

\item We identify the retrieval-reasoning misalignment as a fundamental limitation of existing search agents and analyze its manifestations in unstable search trajectories. We propose R$^2$-Searcher, a novel agentic search framework that dynamically calibrates the retrieval–reasoning boundary, enabling retrieval and reasoning to co-evolve through iterative interactions.

\item We explicitly guide the agent's attention to concise reasoning contexts through query tokens split by pasts of speech, and perform targeted reflection for better search policy. We train R$^2$-Searcher with an end-to-end RL algorithm R$^2$PO, which explores and optimizes the step-level reasoning regions and reflection with process-level signals in tree-based rollouts.

\item We conduct comprehensive and extensive experiments on seven benchmark datasets to evaluate our R$^2$-Searcher. The results demonstrate its superior performance on agentic search tasks, improving the search and reasoning quality simultaneously.
\end{itemize}

\section{Problem Formulation}
\label{problem}
In this section, we will introduce the agentic search task that requires multi-step reasoning and retrieval augmented generation. Given a complex question-answer generation task in the original query $\boldsymbol{q}_o$ that requires multi-step reasoning and retrieval. The agentic search system generates accurate answers by thinking, raising queries, and retrieving documents iteratively.

We formulate the agentic search as a Markov Decision Process (MDP), denoted by $\mathcal{M} := (\mathcal{S}, \mathcal{A}, \mathcal{F}, \mathcal{O}, \mathbb{P}, R, L)$. The state space $\mathcal{S} := \mathcal{V}^*$ represents the set of token sequences accumulated in the search trajectory. The action space $\mathcal{A}$ comprises three subspaces in most existing agentic search methods~\cite{searchr1, autorefine}: (1) Think before raising search queries. (2) Raise new search queries based on the current observations and trajectory (3) Give the final answer based on the trajectory. An action $a_t$ is selected via policy $\pi_\theta: \mathcal{S} \to \Delta(\mathcal{A})$. Tool space $\mathcal{F}$ denotes the set of available retrieval functions; observation space $\mathcal{O}:= \mathcal{V}^*$ contains tokenized results returned by tool execution. The transition kernel $\mathbb{P}: \mathcal{S} \times \mathcal{A} \to \Delta(\mathcal{S})$ is deterministic: after executing $a_t$ and retrieving $\boldsymbol{o}_{t+1}$, the next state can be derived using the transition kernel: $\boldsymbol{s}_{t+1}:= (\boldsymbol{s}_t, \boldsymbol{a}_t, \boldsymbol{o}_t)$. The reward function $R$ is utilized to measure the quality of the final output $(\boldsymbol{s}_T, \boldsymbol{a}_T)$ of the search agent. The integer $L \in \mathbb{N}^+$ represents the maximum number of search steps. Our aim is to maximize the reward expectation of the search trajectory, training the search agent to answer the question accurately:
\begin{equation}
\label{obj}
    \max_{\theta} \mathbb{E}_{\boldsymbol{s}_1 \sim \mu(\cdot),\, (\boldsymbol{s}_T, \boldsymbol{a}_T) \sim (\pi_\theta, \mathbb{P})} \left[ R(\boldsymbol{s}_T, \boldsymbol{a}_T) \right]
\end{equation}
where the expectation is taken from the initial prompt distribution $\boldsymbol{s}_1=(\boldsymbol{I}, \boldsymbol{q}_o)\sim \mu(\cdot)$ and the final output $(\boldsymbol{s}_T, \boldsymbol{a}_T)$ generated in $\mathcal{M}$ under the LLM policy $\pi_{\theta}$. $\boldsymbol{I}$ denotes the initial instruction. 


\section{Methodology}
In this section, we present our \textbf{R$^2$-Searcher}, including the rollout strategies and training algorithms. 
Section~\ref{overview} overviews the framework. Section~\ref{traj} describes trajectory generation with step-wise reasoning regions and retrieval reflection. Section~\ref{training_stage} introduces our training pipeline, including supervised fine-tuning and our end-to-end RL algorithm R$^2$PO.


\subsection{Framework Overview}
\label{overview}
Existing search agents often suffer from two critical limitations. First, as the search progresses, they receive lengthy and semantically complex documents from the retriever, which may cause LLMs to reason on contexts irrelevant to the questions ~\cite{mem1,supo}. Second, when faced with a failed retrieval that does not contain answers, these agents struggle to find effective solutions. Consequently, they tend to propagate errors through the trajectory, becoming trapped in a cycle of incorrect retrieval and flawed reasoning.

To address these challenges, we propose \textbf{R$^2$-Searcher}, shown in Figure \ref{fig:overview}, a novel framework that jointly models step-wise and fine-grained reasoning regions and retrieval reflection to escape the retrieval-reasoning trap. Figure \ref{fig:overview}(a) depicts the trajectory generation process. The agent decomposes the current query into linguistically grounded token groups, which serve as anchors to extract fine-grained factual elements from retrieved documents. These elements form a fine-grained \textbf{reasoning region} that guides the reasoning attention of the agent, which is then used to assess whether the retrieval is sufficient, insufficient, or mismatched. For example, the agent reflects that retrieved texts do not contain who has the top album, so it adds the constraint "Who was the top pop artist" based on the extracted factual elements, emphasizing the target of the question is a "pop artist". In the next iteration, the agent is able to search more accurately based on this constraint and focus on the token group "pop artist" for reasoning on effective information. Based on this diagnosis, the agent generates a \textbf{retrieval reflection} that articulates the deficiency and guides the formulation of a refined next-step query. This step-wise, fine-grained reasoning and reflection can enhance each other, which can be optimized in an evolving pattern in the training stage.

Our training pipeline consists of two stages. First, we perform supervised fine-tuning (SFT) shown in Figure \ref{fig:overview}(b) using carefully curated data to teach the backbone LLM to generate fine-grained reasoning regions and reflection aligned with different retrieval outcomes. Second, we introduce Reasoning-Reflection Guided Policy Optimization (R$^2$PO) in Figure \ref{fig:overview}(c), an end-to-end reinforcement learning algorithm that uses tree-based rollouts to explore the joint action space of step-wise factual extraction, reflection, and search, optimizing the policy using process-level reward signals.


Through this design, R$^2$-Searcher enables the agent to dynamically calibrate the boundaries of the retrieval and reasoning, transforming failed attempts into opportunities for correction, and ultimately achieving more robust and accurate agentic search.

\subsection{Reasoning and Retrieval Boundaries}
\label{boundary_def}


Current agentic search methods suffer from retrieval–reasoning misalignment: they either retrieve irrelevant documents~\cite{incorrect_retrieval} or perform reasoning based on incorrect content, severely degrading the quality of agentic search. To address this issue, we define the \emph{retrieval boundary} and \emph{reasoning boundary}, and leverage these precise definitions to guide LLMs in mitigating misalignment.
Formally, given the $t$-th step query $\boldsymbol{q}_t$, let $\mathcal{E}_{\boldsymbol{q}_t}$ denote the minimal set of factual elements required in the observation space to correctly answer $\boldsymbol{q}_t$. We define the retrieval and reasoning boundary as follows.

\noindent\textbf{Retrieval Boundary}: The retrieval is defined to be \emph{correct} w.r.t.~$\boldsymbol{q}_t$ if the newest observation $\boldsymbol{o}_t$ contains all necessary evidence, i.e.,
\begin{equation}
\label{eq2}
    \forall \boldsymbol{e} \in \mathcal{E}_{\boldsymbol{q}_t}, \quad \boldsymbol{e} \subseteq \boldsymbol{o}_t.
\end{equation}
When proposition~\eqref{eq2} does not hold, we distinguish two failure cases: (i) If $\forall \boldsymbol{e} \in \mathcal{E}_{\boldsymbol{q}_t}, \boldsymbol{e} \not \subseteq \boldsymbol{o}_{t}$, the retrieval is \emph{completely mismatched}. (ii)  If $\exists~\boldsymbol{e} \in \mathcal{E}_{\boldsymbol{q}_t}, \boldsymbol{e} \not \subseteq \boldsymbol{o}_{t}$, but not all do, the retrieval is \emph{insufficient}.

\noindent\textbf{Reasoning Boundary}: The reasoning is \emph{correct} w.r.t.~$\boldsymbol{q}_t$ if the agent’s internal inference over the current state $\boldsymbol{s}_t$ and observation $\boldsymbol{o}_{t-1}$ successfully identifies and utilizes $\mathcal{E}_{\boldsymbol{q}_t}$, i.e.,
\begin{equation}
\label{eq3}
    \mathcal{E}_{\boldsymbol{q}_t} \subseteq \mathcal{K}_t=\{\boldsymbol{m}_t\mid \boldsymbol{a}_{t} \sim \pi_{\boldsymbol{\theta}}(\cdot \mid\boldsymbol{m}_t, \boldsymbol{s_{t}}), \boldsymbol{m}_{t}\subseteq\boldsymbol{o}_{t}\},
\end{equation}
where $\mathcal{K}_t$ represents the set of factual elements $\boldsymbol{m}_t$ from the last observation $\boldsymbol{o}_{t}$, based on which the policy $\pi_{\boldsymbol{\theta}}$ generates the next-step action, including generating the fine-grained reasoning region, explicit reflection and modified search query.
If Proposition~\eqref{eq3} is violated, the reasoning deviates from the correct reasoning boundary. In the following sections, we define $\mathcal{K}_t$ as the correct reasoning region if the proposition~\eqref{eq3} holds.

Critically, correct reasoning requires correct retrieval: if the retrieval is mismatched, the reasoning boundary and the answer must be incorrect (Figure \ref{fig:challenges}a). However, even when retrieval is correct ($\mathcal{E}_{\boldsymbol{q}_t} \subseteq \boldsymbol{o}_{t}$), the reasoning may still fail to cover $\mathcal{E}_{\boldsymbol{q}_t}$ (Figure \ref{fig:challenges}b). This may lead the search trajectory in the wrong direction. 
To tackle these issues, we develop R$^2$-Searcher, a framework that effectively escapes the failed retrieval-reasoning trap by transforming failed retrieval and reasoning attempts into searching for effective information and generating accurate answers, which is shown in Figure \ref{fig:challenges}c. This is achieved through a step-wise query-token-guided reasoning process and a retrieval reflection mechanism.

\subsection{Trajectory Generation}
\label{traj}

\subsubsection{\textbf{Query Tokens Guided Reasoning Region}}
Existing agentic search approaches refine the retrieved information directly based on the semantics of queries. However, when the semantics of the query are complex or vague, these refinement methods might ignore key information to generate answers~\cite{infogain}. To address this challenge, we introduce the decomposition of the query tokens based on part-of-speech. We denote the current $t$-th query as $\boldsymbol{q}_t=[q_{t,1}, q_{t,2}, \dots, q_{t,L_t}]$, where $L_t$ is the number of tokens in the query. Inspired by linguistics, nouns and pronouns typically denote \textit{temporal markers} and \textit{subjects} such as people and places. Verbs express actions that connect the subject to the action. Adjectives and adverbs serve as degree modifiers, which describe the characteristics of subjects and verbs~\cite{part_of_speech, zhang2026memsearch}. Thus, the query tokens can be decomposed into $N_t$ token groups $\boldsymbol{\mathcal{C}}_t=\{\boldsymbol{c}_{t,k} \mid \boldsymbol{c}_{t,k} \subseteq \boldsymbol{q}_t, k\in[1, N_t]\}$, where each token group $\boldsymbol{c}_{t,k}$ is a short phrase or a single word, and belongs to one type of subjects, actions, degree modifiers, and temporal markers. 
Since the query sentence is decomposed based on part-of-speech in a complete manner, all contents relevant to decomposed token groups necessarily include $\boldsymbol{\mathcal{K}}_t$, which satisfies Proposition~\eqref{eq3} and forms an initial correct reasoning region.
These token groups are used to guide the agent's reasoning attention to generate relevant factual elements, which improves the quality of reasoning.

In the trajectory generation process, the token split and the reasoning region generation are performed simultaneously. After that, the 
token group $\boldsymbol{c}_{t,k}$ will be expanded into a single sentence $\boldsymbol{x}_{t,k}$ directly generated by the search agent. The sentences will be gathered to form a reasoning region in \verb|<retrieved_facts>| and \verb|</retrieved_| \verb|facts>|, which is shown in Figure \ref{template}. The generation of a fine-grained reasoning region can be formulated as follows:
\begin{equation}
    \boldsymbol{\mathcal{X}}_{t} = \bigcup\limits_{k=1}^{N_t} (\boldsymbol{c}_{t,k}, \boldsymbol{x}_{t,k}),\ \ \text{where} \ (\boldsymbol{c}_{t,k}, \boldsymbol{x}_{t,k}) \sim\pi_{\boldsymbol{\theta}}(\cdot\mid\boldsymbol{s}_t),
\end{equation}
where $\boldsymbol{\mathcal{X}}_t$ at search step $t$ is the generated reasoning region with query tokens and corresponding factual elements. Using this method, the contents related to the query are fully extracted, forming an explicit reasoning region for the current retrieval, which enables the search agent to focus on the fine-grained information and perform correction retrieval reflection.

\subsubsection{\textbf{Retrieval Reflection Mechanism}}
As introduced in Section \ref{boundary_def}, at each search step, the agent will face a retrieval that is sufficient, insufficient, or completely mismatched, which prevents the agent from seeking effective information and performing correct reasoning. Therefore, our R$^2$-Searcher will analyze the constraints that should be applied to the current query and raise the augmented query for higher-quality retrieval. The retrieval reflection and corresponding new search query can be derived by:
\begin{equation}
    \boldsymbol{\mathcal{R}}_t, \boldsymbol{q}_t \sim \pi_{\boldsymbol{\theta}}(\cdot\mid\boldsymbol{s}_t, \boldsymbol{\mathcal{X}_t}).
\end{equation}

For the three cases of the retrieval boundary discussed in Section \ref{boundary_def}, R$^2$-Searcher is trained to learn reflection abilities \textbf{correspondingly}: (1) If the retrieval is \textbf{sufficient} for answering the current question, the agent will directly generate the answer. (2) If the current retrieval is \textbf{insufficient}, the agent will raise a new search query to seek the missing information, so that it can be supplied to answer the current question. (3) If the current retrieved contents are \textbf{completely mismatched}, the agent will analyze the reason for this failure and add constraints to query for more accurate retrieval, thereby calibrating the retrieval boundary. After the supervised fine-tuning, the agent will follow the three types of reflection patterns, which can be explored and optimized in RL training, leading the search trajectories to better directions.

It is noteworthy that the fine-grained reasoning region and the reflection are the collaborative parts in the trajectory. On one hand, generating a fine-grained reasoning region that enhances the quality of the inference by forcing the agent to focus on the useful contexts for questions, and helps the agent to identify the correctness of the retrieval. On the other hand, the calibrated retrieval boundary leads to more accurate reasoning contents, which form an evolving loop for both inference and training. The generation template for the complete rollout process is shown in Figure \ref{template}.

\begin{figure}[t]
    \centering
    \begin{tcolorbox}[colback=gray!3, colframe=black, title=\textbf{Generation Template of R$^2$-Searcher}]
    \small
    \textbf{Input}: Initial prompt $\boldsymbol{I}$ and input question $\boldsymbol{q}_o$.\\
    \textbf{Generated Trajectory}: \\
    \tagsearch{search} Search Query $\boldsymbol{q}_1$: Where do the ovarian artery and uterine artery anastomose specifically? \tagendsearch{search}\\
    \tagdoc{documents}Doc 1...,
    Doc 2...,
    Doc 3...\tagenddoc{documents}\\
    Reasoning Region $\boldsymbol{\mathcal{X}}_2$: \tagfacts{retrieved\_facts}\\
    \quad - [uterine artery] The uterine artery supplies oxygenated blood... \\
    \quad - [anastomose] The ovarian arteries may anastomose with the uterine artery.\tagendfacts{retrieved\_facts}\\
    Reflection $\boldsymbol{\mathcal{R}}_2$ on insufficient retrieval : \taganal{analysis}The retrieval is insufficient as it correctly identifies... However, the retrieved facts do not mention... Need to search for ...\tagendanal{analysis}.\\
    \textbf{OR} Reflection $\boldsymbol{\mathcal{R}}_2$ on mismatched retrieval: \taganal{analysis}The retrieval is mismatched. Add constraints to query: ...\tagendanal{analysis}.\\
    \tagsearch{search} Search Query $\boldsymbol{q}_2$\tagendsearch{search}\\
    ...\\
    Reflection on sufficient retrieval $\boldsymbol{\mathcal{R}}_T$: \taganal{analysis}The retrieved facts are sufficient and directly answer the question. They clearly state that...\tagendanal{analysis}\\
    \tagans{answer} Correct answer $\boldsymbol{a}$ \tagendans{answer}
    \end{tcolorbox}
    \vspace{-2mm}
    \caption{Generation template for R$^2$-Searcher with query tokens guided reasoning region and reflection mechanism.}
    \label{template}
    \vspace{-5mm}
\end{figure}

\subsection{Training Strategies}
\label{training_stage}
\subsubsection{\textbf{Supervised Fine-tuning}}
As the generation of trajectory introduced in Section \ref{traj} is structured and complex, including flexible cases of reasoning contexts and reflection, it is hard to directly train the LLM with such composite capabilities~\cite{sft_rl}. Therefore, we conduct SFT on the backbone LLM to equip the LLM with the basic ability to extract the factual elements to explicitly model the reasoning region and perform corresponding reflection. 
We curate the SFT data with two categories and 1617 data entries: (1) Given the initial prompt, generate the first search query. (2) Given the initial prompt, the first search query and corresponding retrieved documents, generate factual elements and reflection on sufficient/insufficient/mismatched retrieval. In the training data, there are 179 cases of mismatched search, 212 cases of insufficient search, 455 cases of sufficient search, and 771 examples of thinking and raising the first query only. If the retrieval is mismatched or insufficient, new search queries are also appended to the training data, following the template in Figure \ref{template}.

The data generation process consists of three steps. (1) First, we sample some questions from the training dataset and utilize Qwen3-max~\cite{qwen3} as the teacher model to generate the search query for the first step. (2) Second, we sample another group of training data, and use the SpaCy\footnote{https://spacy.io/} toolkit to decompose the query token groups. (3) Third, after deriving the split query tokens, we use the Qwen3-max to extract factual elements to prepare for the reasoning region and retrieval reflection. After conducting SFT on the backbone LLM, the agent model is trained with basic factual elements, reflection, and multi-turn search ability. 

\subsubsection{\textbf{Reasoning-Reflection Guided Policy Optimization}}
Existing search agents ignore the internal structure of the decision process~\cite{atomsearcher}, which prevents the correct assigning of rewards across retrieval-reasoning boundaries. To address this, we design the reasoning-reflection guided Policy Optimization (R$^2$PO) that leverages tree-based rollouts to effectively explore the action space to derive better reasoning-reflection paths and calibrated boundaries. The detailed algorithm is introduced as follows.

\noindent \textbf{Tree-based Rollouts.}
Traditional search agents restrict their action space to only search and answer, so they are easy to optimize via chain-based rollouts. In contrast, the action space $\mathcal{A}$ of R$^2$-Searcher contains three types of process-level actions: search, factual element extraction, and reflection. Consequently, given the trajectory $\boldsymbol{\tau}_t$ at $t$-th step, the model can hardly explore the reasoning-reflection and search action for each search step using chain-based rollouts~\cite{treegrpo}. To address this challenge, we employ a tree-based rollout structure that expands the reasoning-reflection action space with a shared trajectory prefix, enabling the LLM to learn to calibrate the step-wise reasoning and reflection boundaries with process-level signals. 

Firstly, we generate $M$ independent chain-based trajectories based on the initial prompt $\boldsymbol{s}_1^{(i)}=(\boldsymbol{I}^{(i)}, \boldsymbol{q}_o^{(i)})$:
\begin{equation}
\label{init}
    \mathcal{T} = \big\{ \boldsymbol{\tau}^{(i)}\sim \pi_{\boldsymbol{\theta}}(\cdot\mid\boldsymbol{I}^{(i)}, \boldsymbol{q}_o^{(i)})\big\}_{i=1}^{M},
\end{equation}
where $\mathcal{T}$ represents the initialized tree set with size $M$. 

Then we sample $n$ nodes from each chain-based trajectory $\boldsymbol{\tau}^{(i)}$ for expansion, obtaining $n$ branches with some shared prefixes, which consist of a new tree $\boldsymbol{T}_{\text{new}}^{(i)}$. The set of sampled nodes $\mathcal{N}$ is:
\begin{equation}
    \mathcal{N} = \left\{ 
    \left( \boldsymbol{\tau}_{<t_j}^{(i)}, (\boldsymbol{\mathcal{X}}_{t_j}^{(i)},\, \boldsymbol{\mathcal{R}}_{t_j}^{(i)},\, \boldsymbol{q}_{t_j}^{(i)}), \boldsymbol{o}_{t_j}^{(i)}\right) 
    \;\middle|\; 
    i \in [1,M],\; j \in [1,N]
\right\},
\end{equation}
where $ \left( \boldsymbol{\tau}_{<t_j}^{(i)}, (\boldsymbol{\mathcal{X}}_{t_j}^{(i)},\, \boldsymbol{\mathcal{R}}_{t_j}^{(i)},\, \boldsymbol{q}_{t_j}^{(i)}), \boldsymbol{o}_{t_j}^{(i)}\right)$ is the $j$-th sampled node at step $t_j$ in $i$-th tree and $t_j\in[2, T_i-1], t_j \in \mathbb{N}^+$, $T_i$ is the depth of $i$-th tree. Considering that after SFT, the extracted factual elements, reflection, and query are consistent with each other, we collect them together as an action group to jointly explore the retrieval-reasoning boundary dynamics. 

We use the prefix trajectory of each sampled node $\boldsymbol{\tau}^{(i)}_{<t_j}$ for the tree expansion. Specifically, the suffix trajectory $\boldsymbol{\tau}_{\geq t_j}^{(i)}$ is generated by the prefix and then concatenated to it, forming a new branch for the tree at step $t_j$. After generating the suffixes for the $n$ sampled nodes, the chain $\boldsymbol{\tau}^{(i)}$ is expanded to the new tree $\boldsymbol{T}_{\text{new}}^{(i)}$:

\begin{equation}
    \boldsymbol{T}_{\text{new}}^{(i)}=\big\{\boldsymbol{\tau}_{< t_j}^{(i)} \oplus
    \boldsymbol{\tau}_{\geq t_j}^{(i)} \mid\boldsymbol{\tau}_{\geq t_j}^{(i)} \sim \pi_{\boldsymbol{\theta}}(\cdot \mid \boldsymbol{\tau}^{(i)}_{<t_j})\big\}.
\end{equation}
Finally, we obtain the tree set $\boldsymbol{T}^{(i)}\leftarrow \boldsymbol{T}^{(i)} \cup \boldsymbol{T}_{\text{new}}^{(i)}$ for each prompt. Each prompt contains $M\times N$ rollouts, which forms an intra-tree group for advantage computation.

\noindent\textbf{Training Objective}
Based on the tree-based rollout, we derive the objective of the Reasoning-Reflection Guided Policy Optimization:
\begin{equation}
\label{eq:tree_grpo_loss}
\begin{aligned}
    \mathcal{J}(\boldsymbol{\theta}) = 
\mathbb{E}_{\boldsymbol{s}_1 \sim \mu(\cdot),\, \boldsymbol{\tau} \sim \pi_{\text{old}}(\cdot \mid x)} 
\Bigg[
\frac{1}{G} \sum_{i=1}^{G} 
\frac{1}{|\boldsymbol{\tau}^{(i)}|} \sum_{k=1}^{|\boldsymbol{\tau}^{(i)}|}
\min\Big(
    r_{i,k}(\boldsymbol{\theta}) \cdot \\
    \hat{A}
    (\boldsymbol{\tau}^{(i)}),\, 
    \mathrm{clip}\big(r_{i,k}(\boldsymbol{\theta}),\, 1-\epsilon,\, 1+\epsilon\big) \hat{A}(\boldsymbol{\tau}^{(i)})
\Big)-\beta \mathbb{D}_{\text{KL}}(\pi_{\boldsymbol{\theta}} \parallel \pi_{\text{ref}})
\Bigg],
\end{aligned}
\end{equation}
where $r_{i,k}(\boldsymbol{\theta}) = \frac{\pi_{\boldsymbol{\theta}}(\boldsymbol{a}_{k}^{(i)} \mid \boldsymbol{s}_t^{(i)})}{\pi_{\text{old}}(\boldsymbol{a}_{k}^{i} \mid \boldsymbol{s}_t^{(i)})}$ is the importance ratio for the trajectory $i$ at $k$-th token, and $\boldsymbol{a}_{k}^{(i)}$ represents the generated action token. $\hat{A}(\boldsymbol{\tau}^{(i)}) = \hat{A}_{\text{Intra-tree}}(\boldsymbol{\tau}^{(i)}) + \hat{A}_{\text{Inter-tree}}(\boldsymbol{\tau}^{(i)})$ is the tree-based advantage following the previous work Tree-GRPO~\cite{treegrpo}, where
\begin{equation}
    \hat{A}_{\text{Intra/inter-tree}}(\boldsymbol{\tau}^{(i)}) = \dfrac{R(\boldsymbol{\tau}^{(i)}) - \mathrm{mean}\big( \{ R(\boldsymbol{\tau}^{(j)}) \}_{j=1}^{G_{\text{intra/inter}}} \big)}{\mathrm{std}\big( \{ R(\boldsymbol{\tau}^{(j)}) \}_{j=1}^{G_{\text{intra/inter}}} \big)}.
\end{equation}
$\epsilon > 0$ is the clipping parameter, and $\beta \geq 0$ controls the strength of KL regularization $\mathbb{D}_{\text{KL}}(\pi_{\boldsymbol{\theta}}\parallel  \pi_{\text{ref}})= \mathbb{D}_{\mathrm{KL}} \big( \pi_{\boldsymbol{\theta}}(\boldsymbol{\tau} \mid \boldsymbol{s}_1) \parallel \pi_{\text{ref}}(\boldsymbol{\tau} \mid \boldsymbol{s}_1) \big)$ with respect to a reference policy $\pi_{\text{ref}}$. This loss encourages the policy to improve action sequences that yield process-level advantages from the outcome signals, effectively exploring the dynamic action space with multiple components for the agent training.

\noindent\textbf{Reward Modeling.} The reward function $R(\cdot)$ in our R$^2$-Searcher is designed for training the agent to search and generate accurate answers. Additionally, a high-quality factual element extraction process should also be rewarded. We design a simple rule-based structural reward for our R$^2$PO training:
\begin{equation}
    R(\boldsymbol{\tau})=
    \begin{cases}
        1 + p_{\text{format}}, \quad &\text{if}\ \ \boldsymbol{a}=\boldsymbol{a}_{\text{true}}\\
        0.3 + p_{\text{fact}} + p_{\text{format}} + p_{\text{lazy}}, &\text{if} \ \ \boldsymbol{a} \neq \boldsymbol{a}_{\text{true}},
    \end{cases}
\end{equation}
where $\boldsymbol{a}$ is the generated final answer, $\boldsymbol{a}_{\text{true}}$ is the correct answer. In this reward function, we design three types of penalties to avoid generating low-quality search trajectories: (1) \emph{Penalty on format}: We set $p_{\text{format}}=-0.1$ if the format of trajectory $\boldsymbol{\tau}$ is invalid. (2) \emph{Penalty on factual element extraction}: To better generate informative factual elements and calibrate reasoning boundaries, we set $p_{\text{fact}}=-0.1$ if $\boldsymbol{a}\nsubseteq \bigcup_{t=1}^T\boldsymbol{\mathcal{X}}_t$. (3) \emph{Penalty on lazy search}. We also encourage the search agent to explore multi-turn search to find the correct answer so that the trajectory can be optimized by reasoning-reflection exploration. Therefore, we set the penalty $p_{\text{lazy}}=-0.1$ if $\boldsymbol{a} \neq \boldsymbol{a}_{\text{true}}$ and the search step is only 1. Based on these reward designs, we can stably train R$^2$-Searcher to find the optimal reasoning region and effectively reflect on the retrieval. 

\section{Experiments}
To evaluate the effectiveness of R$^2$-Searcher, we conduct a comprehensive set of experiments. In this section, we first present the experimental setup and then provide an in-depth analysis of the results. Specifically, our experiments are designed to address the following research questions:
\begin{itemize}[leftmargin=*]
    \item \textbf{RQ1:} How does R$^2$-Searcher perform compared to state-of-the-art agentic search models on single- and multi-hop QA datasets?
    \item \textbf{RQ2:} How does R$^2$-Searcher improve the search and reasoning quality compared with state-of-the-art methods?
    \item \textbf{RQ3:} Can R$^2$-Searcher scale its ability to a more powerful backbone model and achieve better performance?
    \item \textbf{RQ4:} How do different components (including framework and reward components) contribute to R$^2$-Searcher?
    \item \textbf{RQ5:} How do key hyperparameters related to agentic search affect the R$^2$-Searcher's performance?
\end{itemize}

\subsection{Datasets and Evaluation Metrics}
We evaluate agentic search performance on seven QA benchmarks involving search and reasoning challenges. Among them, Natural Questions (NQ)~\cite{nq}, TriviaQA~\cite{joshi2017triviaqa}, PopQA~\cite{popqa} are single-hop QA tasks, while HotpotQA~\cite{hotpotqa}, 2WikiMultiHopQA~\cite{2wikimqa}, Musique~\cite{musique}, and Bamboogle~\cite{bamboogle} require multi-hop reasoning. We report Exact Match (EM) and F1 score as evaluation metrics to compare the performance of different search methods.

\begin{table*}[t]
\centering
\caption{Performance(\%) comparison of R$^2$-Searcher versus baseline methods across various QA benchmarks. Bold denotes the best results, and underline denotes the second-best results.}
\label{tab:rq1_accuracy}
\renewcommand{\arraystretch}{1.03}
\resizebox{\linewidth}{!}{
\begin{tabular}{l *{15}{c} cc}
\toprule
\multirow{3}{*}{Methods} 
& \multicolumn{6}{c}{Single-Hop QA} 
& \multicolumn{8}{c}{Multi-Hop QA} 
& \multicolumn{2}{c}{\multirow{2}{*}{Avg.}} \\
\cmidrule(lr){2-7} \cmidrule(lr){8-15}
& \multicolumn{2}{c}{NQ} & \multicolumn{2}{c}{TriviaQA} & \multicolumn{2}{c}{PopQA}
& \multicolumn{2}{c}{HotpotQA} & \multicolumn{2}{c}{2Wiki} & \multicolumn{2}{c}{Musique} & \multicolumn{2}{c}{Bamboogle} &  \\
\cmidrule(lr){2-3} \cmidrule(lr){4-5} \cmidrule(lr){6-7}
\cmidrule(lr){8-9} \cmidrule(lr){10-11} \cmidrule(lr){12-13} \cmidrule(lr){14-15} \cmidrule(lr){16-17}
& EM & F1 & EM & F1 & EM & F1 & EM & F1 & EM & F1 & EM & F1 & EM & F1 & EM & F1 \\
\midrule
Direct Inference & 11.8 & 17.7 & 29.4 & 34.3 & 12.1 & 16.8 & 13.7 & 20.2 & 21.1 & 24.9 & 1.59 & 5.46 & 4.80 & 6.78 & 13.5 & 18.0 \\
COT & 13.7 & 19.1 & 31.2 & 36.3 & 12.8 & 15.9 & 13.6 & 19.7 & 19.4 & 22.9 & 2.51 & 6.95 & 5.60 & 10.1 & 14.0 & 18.7 \\
RAG & 26.0 & 31.7 & 42.4 & 47.8 & 26.2 & 29.1 & 14.2 & 19.3 & 7.88 & 11.0 & 2.91 & 5.75 & 7.20 & 11.7 & 18.1 & 22.3 \\
IRCOT & 20.7 & 30.5 & 38.4 & 48.2 & 27.0 & 36.7 & 13.2 & 19.9 & 7.15 & 13.6 & 2.98 & 7.92 & 12.0 & 21.8 & 17.3 & 25.5 \\
Search-o1 & 22.2 & 32.9 & 42.8 & 50.1 & 29.1 & 35.1 & 15.5 & 22.5 & 7.74 & 12.8 & 3.83 & 9.59 & 12.8 & 21.6 & 19.1 & 23.6 \\
Search-R1 & 32.9 & 41.8 & 52.5 & 59.7 & 35.0 & 40.4 & 30.1 & 39.0 & 30.1 & 35.7 & 10.2 & 15.9 & 25.6 & \underline{35.4} & 31.0 & 38.3 \\
ZeroSearch & 34.5 & 43.5 & 52.8 & 60.3 & 38.8 & 42.6 & 24.4 & 33.2 & 21.3 & 25.9 & 5.63 & 10.6 & 13.8 & 23.8 & 27.3 & 34.3 \\
ReSearch & 40.2 & 48.7 & 57.4 & 65.5 & 41.3 & 44.9 & 30.8 & 41.1 & 27.4 & 33.6 & 8.07 & 15.5 & 16.8 & 25.9 & 31.7 & 39.3\\
AutoRefine & \underline{42.1} & \underline{50.6} & \underline{58.1} & \underline{66.9} & \underline{42.8} & \underline{47.1} & \underline{33.6} & \underline{43.8} & \underline{33.4} & \underline{38.8} & \underline{10.3} & \underline{17.1} & \underline{26.4} & 34.7 & \underline{35.2} & \underline{42.7} \\
\textbf{R$^2$-Searcher} & \textbf{43.2} & \textbf{52.3} & \textbf{60.0} & \textbf{67.5} & \textbf{44.8} & \textbf{49.2} & \textbf{37.0} & \textbf{47.6} & \textbf{36.7} & \textbf{42.6} & \textbf{11.6} & \textbf{19.0} & \textbf{33.6} & \textbf{42.2} & \textbf{38.1} & \textbf{45.8} \\
\midrule
Relative Improve & 2.6\% & 3.4\% & 3.3\% & 0.9\% & 4.6\% & 2.7\% & 10.1\% & 8.7\% & 9.9\% & 9.8\% & 12.6\% & 11.1\% & 27.3\% & 19.2\% & 8.2\% & 7.3\% \\
\bottomrule
\end{tabular}}
\vspace{-2mm}
\end{table*}

\subsection{Baselines}
In this subsection, we introduce the baselines implemented in our experiments, which can be divided into three categories.
\textbf{(1) Generation without retrieval.}
\textbf{Direct Inference}: Use backbone LLM to directly generate answers to the questions.
\textbf{COT}~\cite{cot}: Utilize the chain of thought to directly infer on the questions.
\textbf{(2) Generation with retrieval.}
\textbf{RAG}~\cite{RAG}: Retrieve from external corpus to enhance the generation.
\textbf{IRCOT}~\cite{ircot}: Interleave COT to guide the multi-step retrieval.
\textbf{Search-o1}~\cite{searcho1}: Allow LLMs to iteratively raise queries, refine, and think for deep retrieval.
\textbf{(3) RL-based agentic search methods.}
\textbf{Search-R1}~\cite{searchr1}: Train the model using RL to think and search iteratively.
\textbf{ReSearch}~\cite{research}: A search agent to think based on step-wise results and raise new search.
\textbf{ZeroSearch}~\cite{zerosearch}: Utilize a simulation LLM to interact with the policy model without a search engine.
\textbf{AutoRefine}~\cite{autorefine}: Refine the step-wise retrieved contents for high-quality reasoning and training.

\subsection{Implementation Details}
We employ Qwen2.5-3B~\cite{qwen2}-Base as the backbone of R$^2$-Searcher. We use an E5-based retriever~\cite{e5} as the local retrieval module, and the knowledge source is the 2018 WikiPedia dump~\cite{wikicorpus}. In the basic retrieval setting, we allow up to 3 search turns, with the retriever returning the top-3 documents per query.
For SFT, we use Qwen3-Max~\cite{qwen3} as the teacher model to generate a training dataset comprising 1,617 examples, and train the backbone model for 500 steps.
In RL training, we combine the training sets from HotpotQA and NQ. The training configuration includes a batch size of 96, a PPO mini-batch size of 48, and 200 training steps. During rollouts, we sample responses with a temperature of 1.0. Retrieved documents for each query are concatenated and truncated to 768 tokens. The actor model is optimized with a learning rate of $10^{-6}$, a warm-up ratio of 0.285, and a KL divergence loss coefficient of $\beta=0.001$.
For tree-based rollouts, we construct $M = 2$ trees and sample $n = 3$ nodes from each tree for expansion, while for other RL-based methods, the group size is set as 6.
During evaluation, we use the official test split when available; otherwise, we evaluate on the dev split. 
For RL-based baselines, we use the Qwen2.5-3B-Base as the backbone, while for baselines without training, Qwen2.5-3B-Instruct is adopted to better follow the initial instructions. 

\begin{figure*}[t]
    \centering
    \begin{subfigure}[t]{0.24\linewidth}
        \centering
        \includegraphics[height=3.04cm]{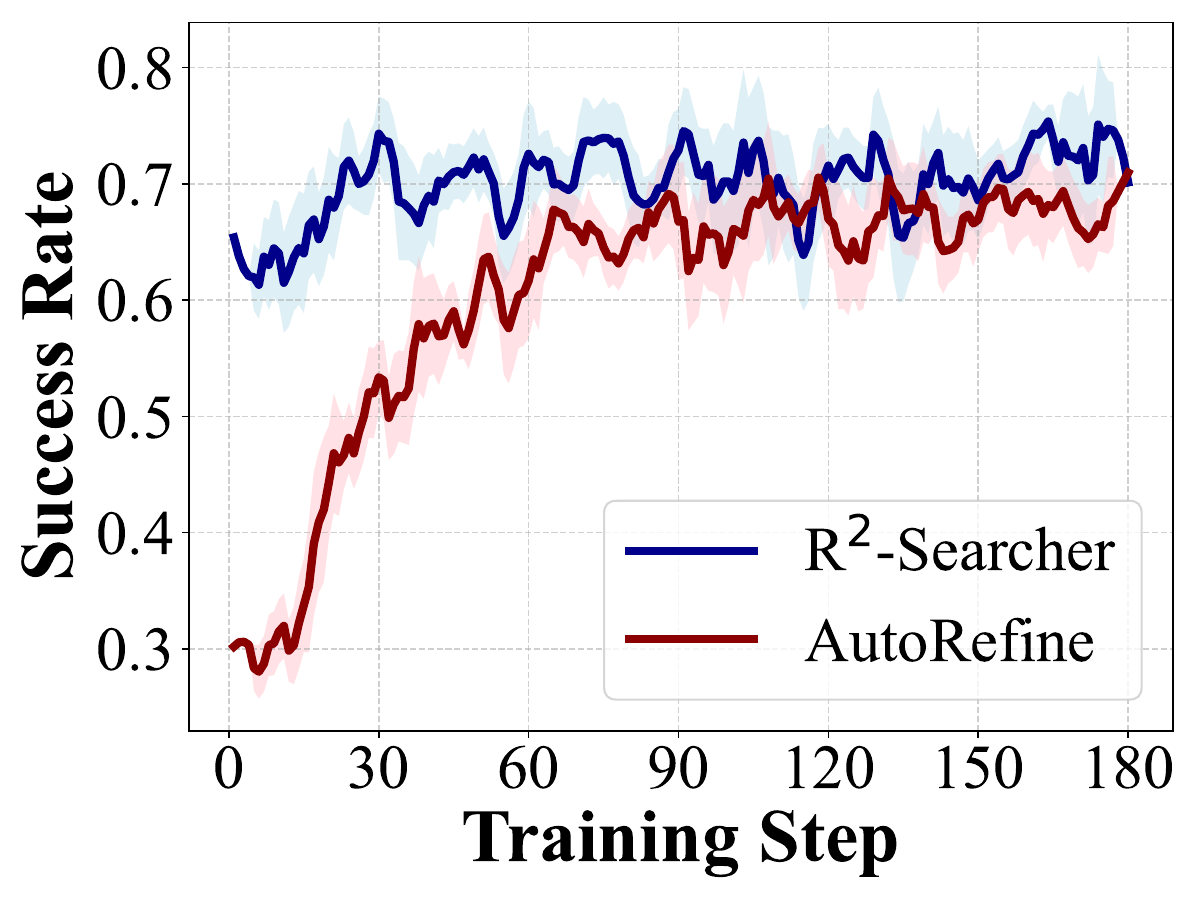} %
        \caption{Search Quality (Train).}
        \label{fig:search}
    \end{subfigure}
    \hspace{1pt}
    \begin{subfigure}[t]{0.24\linewidth}
        \centering
        \includegraphics[height=3.04cm]{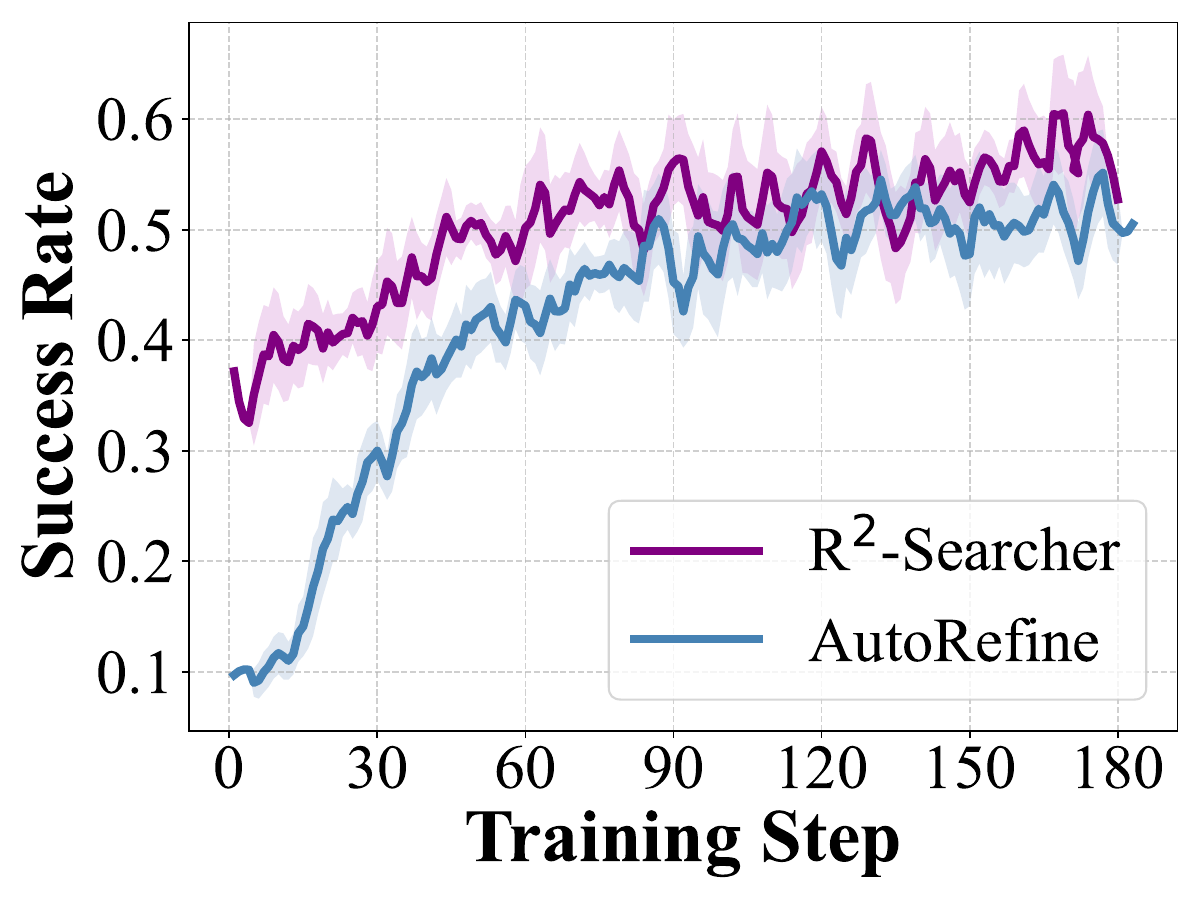}
        \caption{Reasoning Quality (Train).}
        \label{fig:reason}
    \end{subfigure}
    \hspace{1pt}
    \begin{subfigure}[t]{0.24\linewidth}
        \centering
        \includegraphics[height=3.04cm]{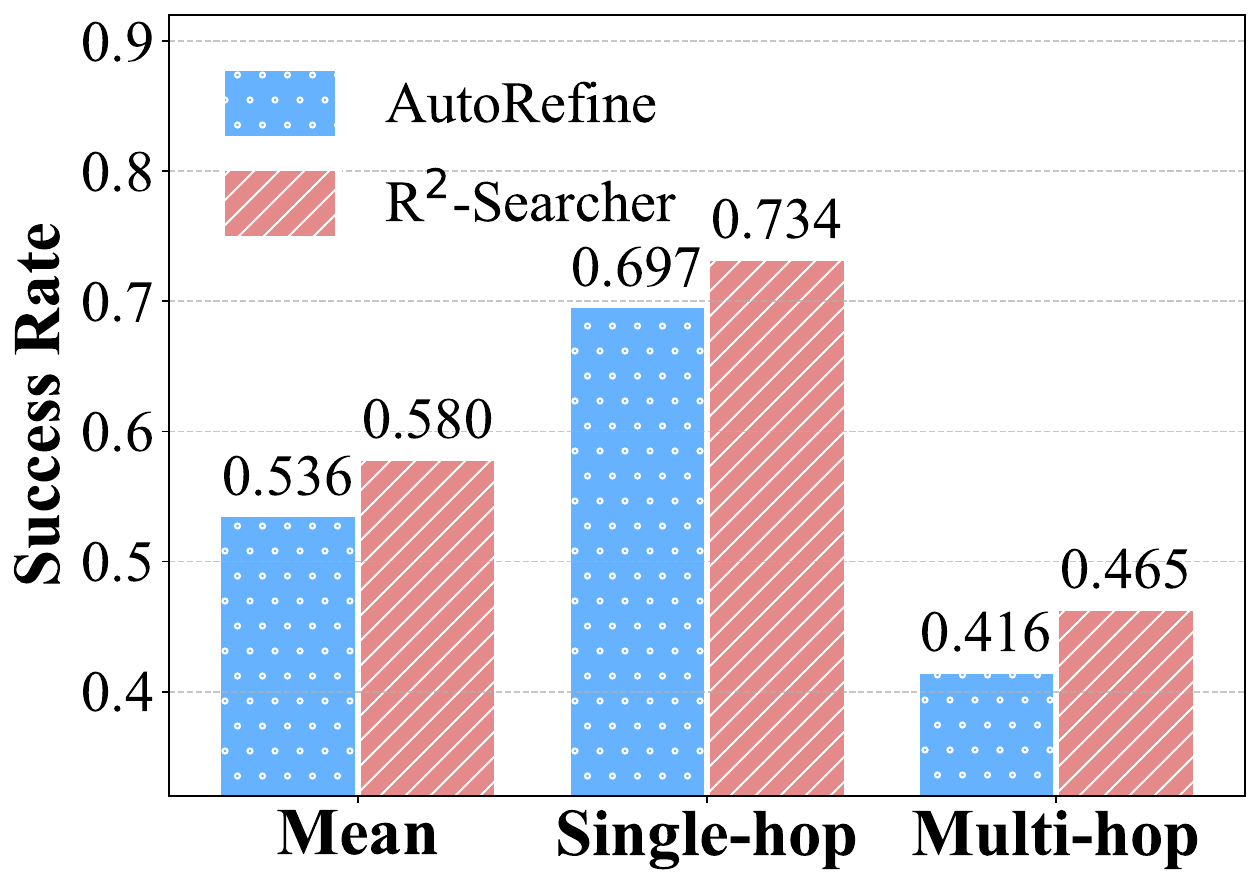}
        \caption{Search Quality (Test).}
        \label{fig:search_test}
    \end{subfigure}
    \hspace{1pt}
    \begin{subfigure}[t]{0.24\linewidth}
        \centering
        \includegraphics[height=3.04cm]{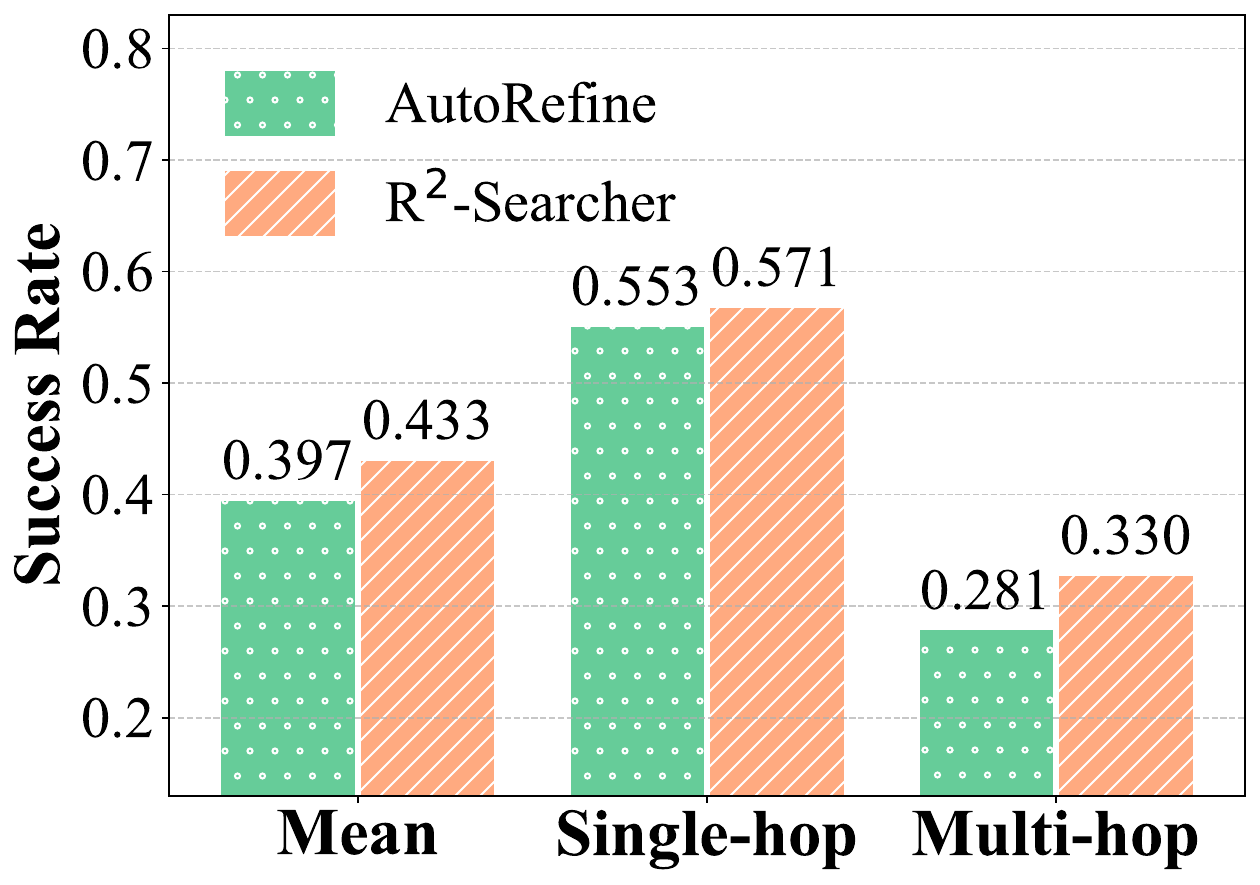}
        \caption{Reasoning Quality (Test).}
        \label{fig:reason_test}
    \end{subfigure}
    \vspace{-3mm}
    \caption{Training Dynamics and Test Performance of AutoRefine and R$^2$-Searcher on Search and Reasoning Quality.}
    \label{training}
    \vspace{-2mm}
\end{figure*}

\subsection{Overall Performance Comparison (RQ1)}
In this subsection, we compare R$^2$-Searcher with state-of-the-art baselines to evaluate its overall effectiveness. The results of answer accuracy are shown in Table \ref{tab:rq1_accuracy}, which presents a comprehensive comparison of R$^2$-Searcher against state-of-the-art baselines across seven benchmarks, reporting both EM and F1 scores.

R$^2$-Searcher consistently achieves the best overall performance across all datasets, and outperforms the previous strongest baseline, AutoRefine, by an average of \textbf{8.2\%} in EM and \textbf{7.3\%} in F1.
Notably, its gains are most significant on multi-hop datasets: on HotpotQA, 2Wiki, and Musique, it improves the EM and F1 score ranging from \textbf{8.7\%} to \textbf{12.6\%}; on Bamboogle, it improves EM by \textbf{27.3\%} and F1 by \textbf{19.2\%}. These results demonstrate that R$^2$-Searcher exhibits strong generalization and robustness in complex, multi-step search and reasoning tasks. This advantage arises from our two key innovations: (1) the explicit \textit{query-token-guided reasoning region} that dynamically identifies and focuses on critical factual elements and (2) the \textit{step-wise reflection mechanism} that enables the agent to iteratively refine its retrieval strategy. Together, these components allow R$^2$-Searcher to navigate search and reasoning trajectories more effectively, yielding higher-quality answers from explicit reasoning regions.

In contrast, existing approaches without training on retrieval tasks exhibit clear limitations. Direct Inference and COT perform poorly across the benchmarks, as they rely solely on internal parametric knowledge and cannot access external evidence. Vanilla RAG shows moderate improvements, but still struggles with multi-hop reasoning due to its inability to decompose complex questions into iterative retrieval steps. Multi-step RAG variants without RL training, including IRCOT and Search-o1, offer only marginal gains, as their backbone LLMs lack training for adaptive retrieval and reasoning, leading to ineffective search trajectories.

Additionally, RL-based agents, including Search-R1, ZeroSearch, ReSearch, and AutoRefine, significantly outperform non-RL methods, demonstrating the effectiveness of learning through interaction. Among them, AutoRefine achieves the strongest prior results (EM = 35.2\%, F1 = 42.7\%), attributed to its step-wise refinement mechanism that enhances reasoning quality. However, AutoRefine's refinement lacks explicit and fine-grained query-aware guidance, and it struggles to deal with mismatched retrieval. In contrast, R$^2$-Searcher generates a concise and accurate \textit{reasoning contexts} based on query tokens, enabling targeted reflection and retrieval guidance. This mutual reinforcement between reasoning focus and \textit{reflective retrieval} allows R$^2$-Searcher to find compact a reasoning region in the contexts with a more effective search path. As a result, R$^2$-Searcher achieves the highest scores across all benchmarks, confirming its effectiveness in handling complex, information-seeking tasks.
Our R$^2$-Search with SFT does not inject additional knowledge from the teacher model, and the detailed analysis in Appendix~\ref{teacher_knowledge}.
In addition, we analyze the efficiency of R$^2$-Searcher in Appendix~\ref{efficiency} to show it has strong applicability like other agentic search methods.

In summary, R$^2$-Searcher is a powerful LLM search agent that integrates explicit reasoning regions with iterative reflection, enabling accurate and interpretable multi-hop retrieval and reasoning.

\begin{table*}[t]
\centering
\caption{Model Performance (\%) on Different LLM Backbones with Different Capabilities.}
\vspace{-2mm}
\label{tab:backbone}
\renewcommand{\arraystretch}{0.9}
\resizebox{\linewidth}{!}{
\begin{tabular}{l *{15}{c} cc}
\toprule
\multirow{3}{*}{Methods} 
& \multicolumn{6}{c}{Single-Hop QA} 
& \multicolumn{8}{c}{Multi-Hop QA} 
& \multicolumn{2}{c}{\multirow{2}{*}{Avg.}} \\
\cmidrule(lr){2-7} \cmidrule(lr){8-15}
& \multicolumn{2}{c}{NQ} & \multicolumn{2}{c}{TriviaQA} & \multicolumn{2}{c}{PopQA}
& \multicolumn{2}{c}{HotpotQA} & \multicolumn{2}{c}{2Wiki} & \multicolumn{2}{c}{Musique} & \multicolumn{2}{c}{Bamboogle} & \\
\cmidrule(lr){2-3} \cmidrule(lr){4-5} \cmidrule(lr){6-7}
\cmidrule(lr){8-9} \cmidrule(lr){10-11} \cmidrule(lr){12-13} \cmidrule(lr){14-15} \cmidrule(lr){16-17}
& EM & F1 & EM & F1 & EM & F1 & EM & F1 & EM & F1 & EM & F1 & EM & F1 & EM & F1 \\
\midrule
AutoRefine (Qwen2.5-3B) & 42.1 & 50.6 & 58.1 & 66.9 & 42.8 & 47.1 & 33.6 & 43.8 & 33.4 & 38.8 & 10.3 & 17.1 & 26.4 & 34.7 & 35.2 & 42.7 \\
\textbf{R$^2$-Searcher} (Qwen2.5-3B) & \textbf{43.2} & \textbf{52.3} & \textbf{60.0} & \textbf{67.5} & \textbf{44.8} & \textbf{49.2} & \textbf{37.0} & \textbf{47.6} & \textbf{36.7} & \textbf{42.6} & \textbf{11.6} & \textbf{19.0} & \textbf{33.6} & \textbf{42.2} & \textbf{38.1} & \textbf{45.8} \\
\midrule
AutoRefine (Qwen3-4B) & 43.9 & 53.8 & 61.8 & 69.3 & 42.9 & 48.7 & 40.1 & 51.4 & 33.8 & 40.7 & 15.7 & 23.5 & 43.2 & 53.0 & 40.2 & 48.6 \\
\textbf{R$^2$-Searcher} (Qwen3-4B) & \textbf{45.4} & \textbf{54.5} & \textbf{62.2} & \textbf{70.3} & \textbf{47.5} & \textbf{52.1} & \textbf{42.4} & \textbf{53.8} & \textbf{42.2} & \textbf{48.1} & \textbf{15.8} & \textbf{23.6} & \textbf{44.8} & \textbf{55.2} & \textbf{42.8} & \textbf{51.0} \\
\bottomrule
\end{tabular}}
\vspace{-2mm}
\end{table*}

\subsection{Search and Reasoning Quality (RQ2)}

    

In this subsection, we evaluate the training dynamics of R$^2$-Searcher and baseline methods. We adopt \textit{retrieval success rate} and \textit{reasoning success rate} to measure retrieval and reasoning quality, respectively. Across this and subsequent sections, a retrieval step is considered successful if the retrieved documents contain the ground-truth answer. A reasoning step is successful if the answer appears within either the \texttt{\textless refine\textgreater} or \texttt{\textless retrieved\_facts\textgreater} segments and the final predicted answer achieves an F1 score of at least 10.0\%. We adopt this low F1 threshold to account for partial but semantically meaningful intermediate reasoning outputs during training.

As shown in Figure~\ref{fig:reason}, our R$^2$-Searcher demonstrates superior training dynamics compared to AutoRefine. In Figure~\ref{fig:search}, R$^2$-Searcher achieves a consistently higher search success rate throughout training, reaching approximately 0.75 by step 180, while AutoRefine reaches about 0.7. This observation indicates its ability to more effectively explore the action space and calibrate retrieval boundaries. Additionally, R$^2$-Searcher exhibits notably stronger performance from the beginning of training compared to AutoRefine. This early advantage stems directly from our SFT phase, which equips the model with foundational capabilities to generate effective search queries and retrieval reflections.

Similarly, in Figure~\ref{fig:reason}, R$^2$-Searcher exhibits faster convergence and higher reasoning success rates: it reaches around 0.6 before step 180, while AutoRefine’s only reaches around 0.52, representing a significant improvement. These results reflect its capability to dynamically refine reasoning regions, thereby avoiding the retrieval-reasoning trap that often plagues baseline methods.

On the test set shown in Figure~\ref{fig:search_test} and Figure~\ref{fig:reason_test}, R$^2$-Searcher significantly outperforms AutoRefine across both search and reasoning quality. Specifically, it achieves \textbf{0.580} in search quality (improving upon AutoRefine by \textbf{8.2\%}) and \textbf{0.433} in reasoning quality (improving upon AutoRefine by \textbf{8.3\%}). Moreover, the gains are most significant in Multi-hop tasks, improving the search and reasoning quality by \textbf{12.0\%} and \textbf{17.4\%} respectively. These observations confirm that R$^2$-Searcher’s structured trajectory generation, including query-token guided reasoning region extraction and retrieval reflection, results in higher-quality search and more accurate answers.

\begin{figure}[t]
    \centering
    \begin{subfigure}[t]{0.48\linewidth}
        \centering
        \includegraphics[height=3cm]{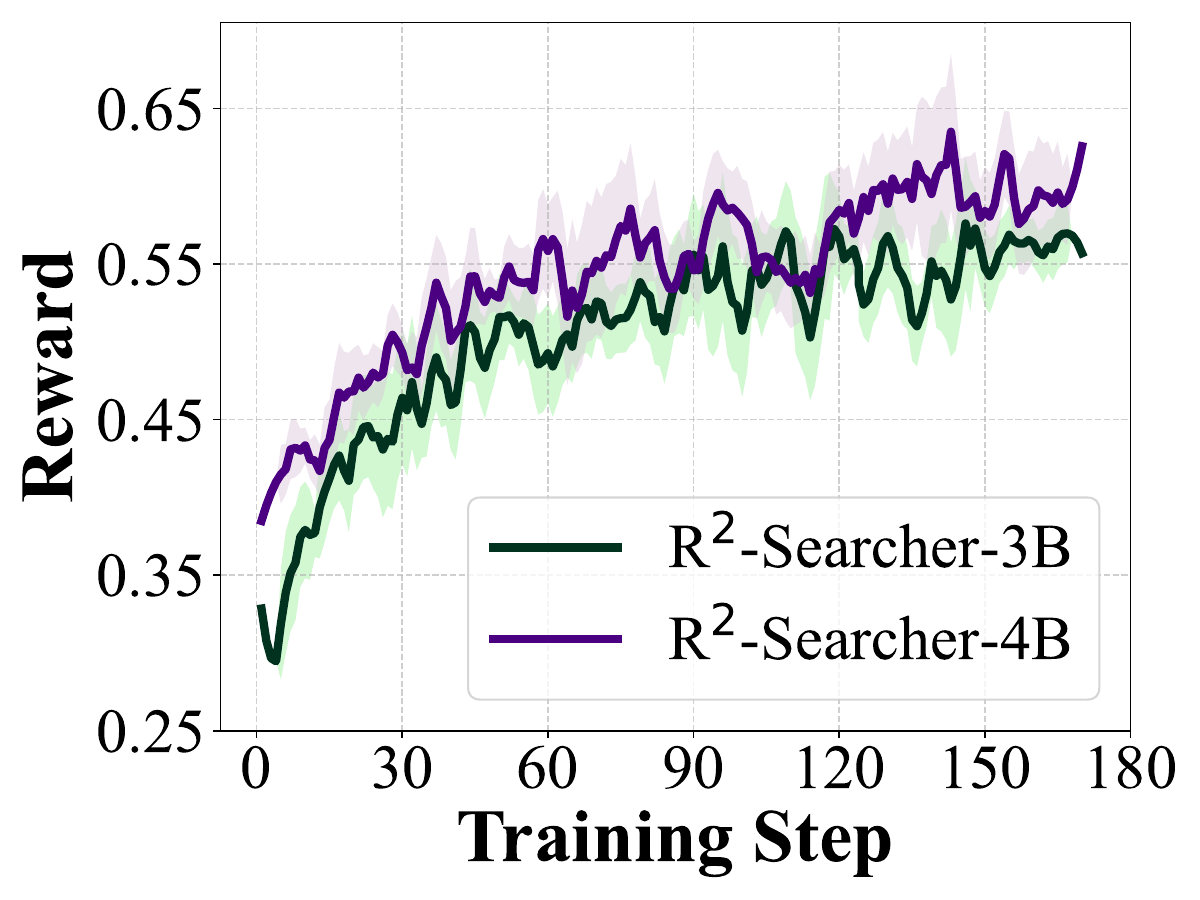} %
        \caption{Training Reward Curves of R$^2$-Searcher-3B and 4B.}
        \label{fig:scale}
    \end{subfigure}
    \hspace{1pt}
    \begin{subfigure}[t]{0.48\linewidth}
        \centering
        \includegraphics[height=3cm]{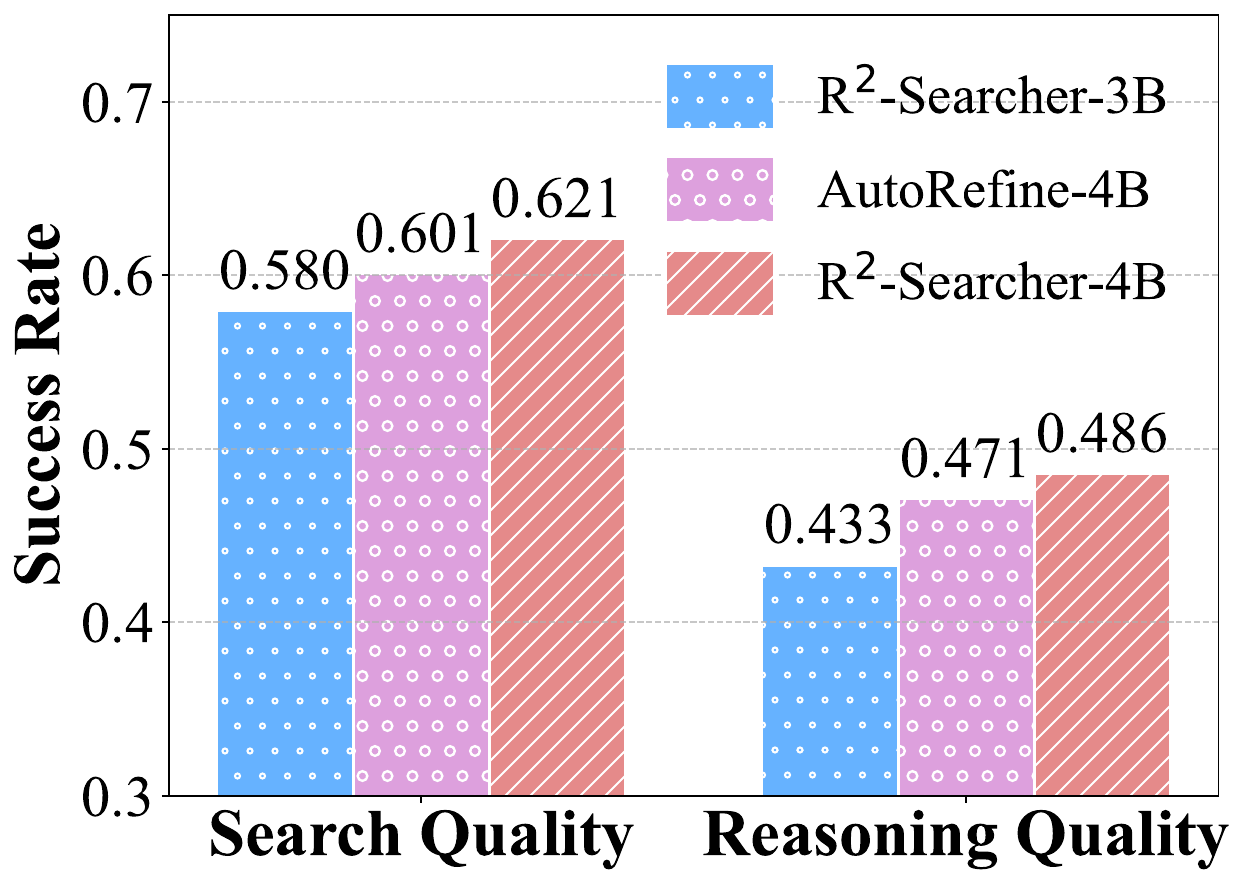}
        \caption{Comparison of Search and Reasoning Quality.}
        \label{fig:4btest}
    \end{subfigure}
    \vspace{-2mm}
    \caption{Training Dynamics of R$^2$-Searcher-3B \& 4B and Their Search \& Reasoning Quality.}
    \label{fig:scale_training}
    \vspace{-5mm}
\end{figure}

\subsection{Scalability of R$^2$-Searcher (RQ3)}
We further conduct experiments to investigate how R$^2$-Searcher scales to a more powerful LLM backbone, Qwen3-4B~\cite{qwen3}, which exhibits performance comparable to Qwen2.5-7B.

Our R$^2$-Searcher consistently delivers significant performance gains across both smaller (Qwen2.5-3B) and larger (Qwen3-4B) LLM backbones, demonstrating its strong scalability and adaptability to more capable foundation models. 
When scaled to Qwen3-4B, R$^2$-Searcher further improves to 42.8\% EM and 51.0\% F1, again outperforming AutoRefine on all sub-datasets, notably improving EM by \textbf{24.9\%} on 2Wiki. The results indicate that R$^2$-Searcher consistently maintains its advantage in enhancing multi-hop search performance.
Moreover, as shown in Figure~\ref{fig:scale}, R$^2$-Searcher-4B consistently achieves higher rewards than its 3B version during RL training and significantly enhances the search and reasoning quality shown in Figure~\ref{fig:4btest}. It also outperforms AutoRefine-4B in both success rate of search and reasoning, yielding relative improvements of \textbf{3.3\%} and \textbf{3.2\%}, respectively. R$^2$-Searcher’s core mechanisms, fine-grained reasoning region modeling, and retrieval reflection, are not only effective for small models but also can effectively leverage the capacity of stronger backbones. The consistent EM and F1 gains across all datasets, especially in challenging multi-hop benchmarks like HotpotQA and 2Wiki, validate that our framework is robustly extensible and capable of unlocking higher-order reasoning and search precision as the backbone’s capacity increases.

\begin{table}
		\caption{\textbf{Ablation study for Components of R$^2$-Searcher.}}
        \vspace{-2mm}
		\label{tab:ablation}
        \resizebox{0.98\linewidth}{!}{
        \renewcommand{\arraystretch}{0.92}
		\begin{tabular}{ccccc}
			\toprule
			Components & EM & F1 & Search & Reasoning \\
			\midrule
            \rowcolor{gray!10} \textbf{\textit{Framework Components}} & & & & \\
            w/o $R^2$PO (SFT only) & 16.2 & 20.5 & 39.2 & 20.5\\
            w/o tree-based rollout & 36.5 & 43.7 & 53.5 & 40.1\\
            w/o Reasoning Region & 34.7 & 41.2 & 52.4 & - \\
            w/o Reflection & 35.7 & 42.6 & 51.2 & 37.1\\
            \midrule
            \rowcolor{gray!10} \textit{\textbf{Reward Components}} & & & & \\
            w/o $p_{\text{fact}}$ & 36.3 & 43.1 & 53.8 & 39.6\\
            w/o $p_{\text{lazy}}$ & 33.2 & 39.8 & 47.9 & 35.1\\
            \midrule
            R$^2$-Searcher & \textbf{38.1} & \textbf{45.8} & \textbf{58.0} & \textbf{43.3}\\
			\bottomrule
		\end{tabular}}
        \vspace{-5mm}
	\end{table}
    
\subsection{Ablation Study (RQ4)}
To further investigate the effectiveness of our R$^2$-Searcher, we conduct an ablation study on the \textbf{framework components} and \textbf{reward components}, which are essential parts for trajectory generation and RL training, respectively. For framework components, we ablate the R$^2$PO, tree-based rollout, query token guided reasoning region, and retrieval reflection mechanism. For ablation on reward components, we remove the essential penalties, including $p_{\text{fact}}$ and $p_{\text{lazy}}$. The results of the ablation study are shown in Table \ref{tab:ablation}.

\noindent\textbf{Framework Components.}
We first remove the R$^2$PO stage, retaining only SFT. The resulting model exhibits strong search capability but significantly weaker reasoning performance compared to the full R$^2$-Searcher, indicating that RL is essential for aligning reasoning and reflection with search objectives. In addition, without the tree-based rollout, the RL algorithm cannot fully explore the calibration of reasoning and reflection, leading to a performance drop.
When either the query-token-guided reasoning region or the reflection mechanism is removed, both search and reasoning performance decrease by 6\%-7\% (Reasoning quality cannot be calculated w/o reasoning region). Without these components, the model cannot establish the mutual reinforcement between reasoning and reflection to produce high-quality trajectories.
Removing the reasoning region forces the model to summarize retrieved content during reflection, yielding limited but non-negligible search performance. Removing the reflection mechanism still preserves moderate search ability, as the model can leverage the reasoning region to implicitly assess known information. 
We do not ablate the SFT stage, as directly training the base model using our trajectory template without SFT leads to training instability.
The complex action space, combined with the model’s weak initial capabilities, leads to invalid 
reasoning regions and reflection contents.


\noindent \textbf{Reward Components.}
Ablating the penalty for factual elements extraction $p_{\text{fact}}$ leads to a significant performance drop: on multi-hop datasets, EM and F1 drop by 2.2 and 2.7 percentage points, respectively. Without this term, the model’s reasoning is no longer constrained by factual grounding during training; consequently, in tree-based rollouts, the process reward is not properly attributed to the current reasoning step, resulting in low-quality factual elements. Similarly, after removing the lazy-search penalty $p_{\text{lazy}}$, the agent may maximize immediate reward through single-step retrieval without engaging in deeper reasoning or reflection, which substantially degrades the performance. Without $p_{\text{lazy}}$, EM and F1 drop by 4.9 and 6.0 points, respectively. This confirms that both reward components are essential for guiding the agent toward thorough, factually grounded, and multi-step search behaviors.

\subsection{Parameter Analysis (RQ5)}
In this subsection, we analyze our R$^2$-Searcher-3B by varying two key search parameters: the maximum number of retrieval rounds $L$ and the number of top-$k$ documents returned. When studying the effect of one parameter, the other is held at its default value.

\begin{figure}[t]
    \centering
    \captionsetup[subfigure]{skip=1pt} 
    
    \begin{subfigure}[b]{0.82\linewidth}
        \centering
        \includegraphics[width=\linewidth]{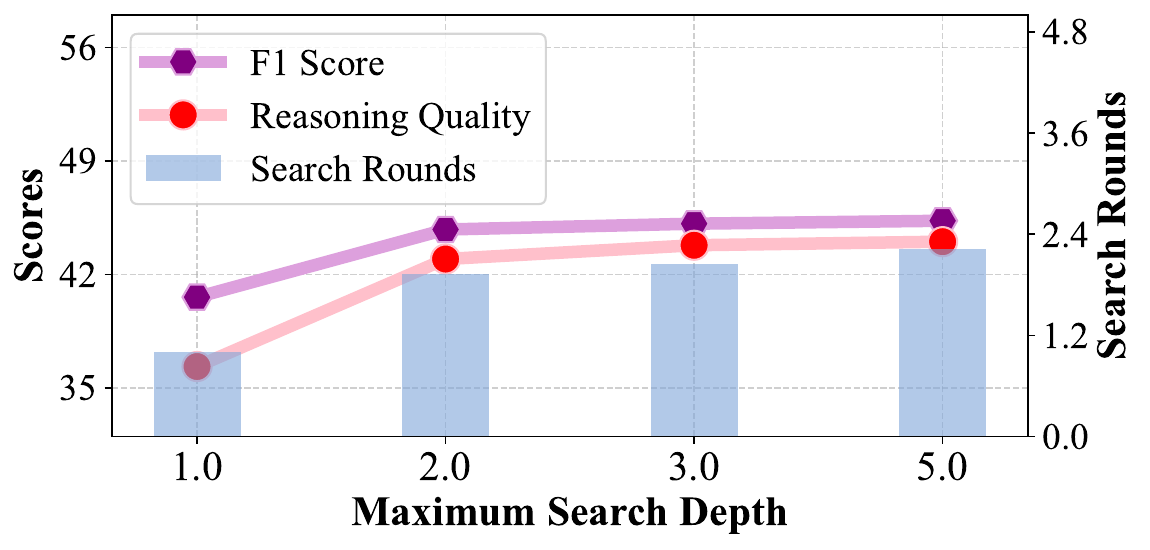}
        \caption{Performance on Search Depth.}
        \label{fig:depth}
    \end{subfigure}
    
    
    \begin{subfigure}[b]{0.82\linewidth}
        \centering
        \includegraphics[width=\linewidth]{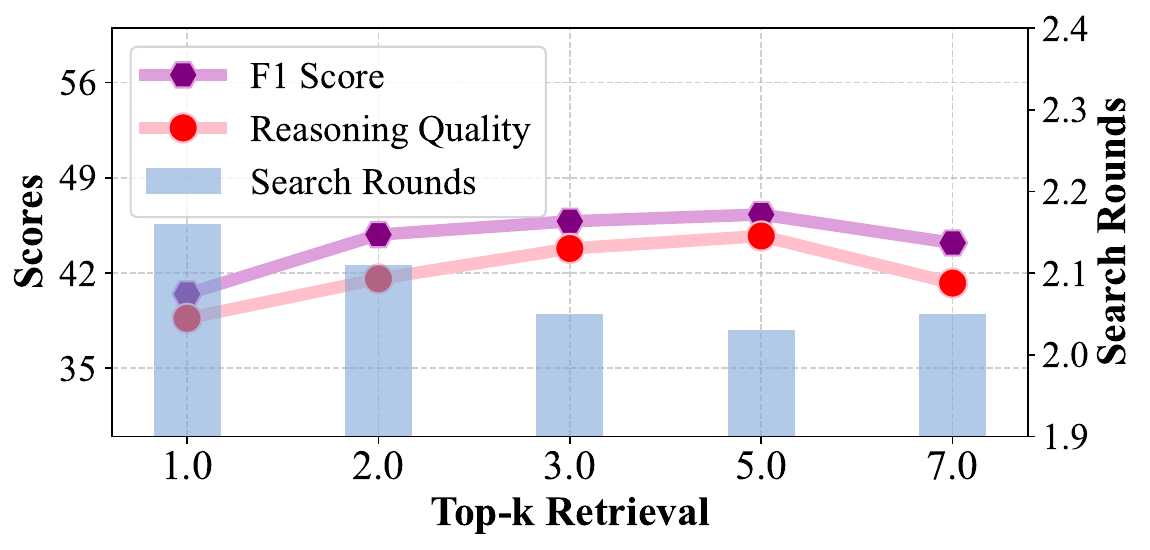}
        \caption{Performance on Top-$k$ Retrieval}
        \label{fig:topk1}
    \end{subfigure}

    \vspace{-3mm}
    \caption{Parameter Analysis of R$^2$-Searcher.}
    \label{topk}
    \vspace{-5mm}
\end{figure}

\noindent\textbf{Search Depth.}
As shown in Figure~\ref{fig:depth}, increasing the maximum retrieval rounds encourages the model to automatically extend its search depth when tackling harder questions, as the search agent has more trials to explore insufficient or mismatched retrieval. However, this increase is modest, as a substantial fraction of questions can already be resolved within two rounds. Moreover, while both retrieval and generation quality improve with more search rounds, gains diminish beyond $L=5$, likely due to inherent limits in model capacity and retrieval quality. These results demonstrate that R$^2$-Searcher has the ability to flexibly raise search queries based on fine-grained reasoning region and retrieval reflection.

\noindent\textbf{Top-$k$ Retrieval.}
Figure~\ref{fig:topk1} illustrates how model performance and search behavior vary with the number of retrieved documents $k$. As $k$ increases, the model gains access to richer information, leading to steady performance improvements. From $k=1$ to 5, the reasoning success rate and the F1 score show an absolute increase of about 5\%. When $k$ becomes too large (e.g., $k=7$), the accumulated context across multiple retrieval rounds grows excessively long, causing attention dilution and a performance drop. 
Besides, the number of search rounds increases when the number of documents is reduced from 3 to 1, leading to information loss and forcing the agent to execute more search turns. This observation shows our R$^2$-Searcher can seek information when retrieval is insufficient or mismatched.

\section{Related Works}

\noindent \textbf{Agentic Retrieval-Augmented Generation.} 
Retrieval-Augmented Generation (RAG)~\cite{RAG}, leveraging external knowledge bases to enhance generation quality, has become widely adopted. Early approaches like Query2doc~\cite{query2doc} and  CBR-RAG~\cite{cbr-rag}, enrich queries with pseudo-documents and relevant cases, improving the retrieval quality. GraphRAG~\cite{graphrag}, HippoRAG~\cite{hipporag}, and HiRAG~\cite{HiRAG} employ graph or hierarchical structures to organize knowledge, boosting reasoning performance.
However, their constrained retrieval scope often fails on complex multi-hop tasks~\cite{amber, amem}. To overcome this, Agentic RAG methods such as IRCOT~\cite{ircot}, Adaptive-RAG~\cite{adatptiverag} and Search-o1~\cite{searcho1}, adopt iterative retrieval strategies to seek answers based on well-designed instructions. Yet the LLM backbones of these methods are not pretrained for interacting with search engines, leading to suboptimal performance~\cite{searchr1, research}. To address this challenge, Search-R1~\cite{searchr1} utilizes end-to-end RL to optimize LLM reasoning with multi-turn interactions. Atom-searcher~\cite{atomsearcher} and Interact-RAG~\cite{hui2025interact} train the model to decompose the thinking process in an explicit pattern. Other variants of Search-R1 improve the reasoning process with memory management~\cite{memsearcher, mem1} and refinement~\cite{autorefine, reasonrag}. ParetoPO~\cite{li2026paretopo} balances the efficiency of tool utilization and the reasoning for the agentic search. Uniquely, our method equips the RL-based search agent with more fine-grained reasoning capability and reflection on the step-wise search.


\noindent \textbf{Context Management.}
Context management has become increasingly popular in LLM agents, as contexts can be managed by the LLM itself to enhance the reasoning in an evolving pattern. Existing context management strategies can be divided into two categories, including history-based~\cite{mem1, supo} and step-wise management~\cite{deepretrieval, autorefine}.  History-based management like SUPO~\cite{supo}, Mem1~\cite{mem1}, and MemAgent~\cite{yu2025memagent} summarizes useful parts of the history and integrates them into a memory state, thereby improving the inference performance. Moreover, some previous works store the contexts in a structural representation, including knowledge graphs~\cite{graphragr1, zep}, atomic notes~\cite{amem}, and hierarchical knowledge~\cite{hu2025hiagent, gmemroy, sun2025hierarchical}. For step-wise context management, the agent refines the contexts in each reasoning step, based on which the subsequent reasoning can be optimized~\cite{markovian}. For example, AutoRefine~\cite{autorefine} trains the LLM to refine and think after obtaining retrieved contents, aiming to enhance the next-step retrieval. ReasonRAG~\cite{reasonrag} and DeepRetrieval~\cite{deepretrieval} refine the generation process through RL training. However, these step-wise refinements fail to integrate fine-grained evidence, limiting the refinement quality. In this paper, R$^2$-Searcher focuses on step-wise explicit reasoning region modeling and reflection mechanism, improving the search and reasoning quality of LLM search agents.

\section{Conclusion}
In this paper, we propose R$^2$-Searcher, a novel LLM search agent that effectively calibrates retrieval and reasoning boundaries through query-token guided reasoning region modeling and a retrieval reflection mechanism. Our work addresses the key challenges of shifted retrieval and reasoning boundaries by: (1) explicitly modeling a fine-grained query-token guided reasoning region, (2) generating a step-wise reflection mechanism according to the retrieved contents, and (3) training the LLM to explore the retrieval-reasoning boundaries with fine-grained reasoning region, reflection, and search queries. Extensive experiments demonstrate that our R$^2$-Searcher outperforms state-of-the-art search agent models, with high search and reasoning quality. Our R$^2$-Searcher introduces a novel approach to collaborative reasoning and reflection, establishing an evolving path toward high-performance agentic search.
\appendix
\begin{appendices}
\section{Clarification for Teacher Knowledge}
\label{teacher_knowledge}
Many existing works leverage LLMs to generate high-quality knowledge, data or trajectories to enhance the performance of more lightweight models using post trainig~\cite{atomsearcher, wang2025embedding, guo2025deepseek, hu2022lora}. However, for fair comparison in the experiment, the effects of knowledge from teacher LLM should be clarified. 
In this section, we demonstrate that the knowledge of the teacher model (Qwen3-max) was not injected into the student model (Qwen2.5-3B-Base). 
We compare the performance of the supervised fine-tuned model based on the teacher trajectories with Naive RAG and R$^2$-Searcher.

\begin{table}
		\caption{\textbf{Performance (\%) comparison among SFT, Naive RAG and R$^2$-Searcher. The best results are bolded.}}
        \vspace{-2mm}
		\label{tab:teacher}
        \renewcommand{\arraystretch}{1}
		\begin{tabular}{ccccc}
			\toprule
			Methods & EM & F1 & Search & Reasoning \\
			\midrule
            SFT only & 16.2 & 20.5 & \underline{39.2} & \underline{20.5}\\
            Naive RAG & \underline{18.1} & \underline{22.3} & 30.6 & 9.90\\
            R$^2$-Searcher & \textbf{38.1} & \textbf{45.8} & \textbf{58.0} & \textbf{43.3}\\
			\bottomrule
		\end{tabular}
        \vspace{-2mm}
\end{table}
The results in Table~\ref{tab:teacher} show that the model fine-tuned by the trajectories curated by Qwen3-max does not improve the overall quality of its answer. After applying SFT, the EM and F1 scores decrease by 1.9 and 1.8 points, respectively, indicating that no additional knowledge was injected into the base model of R$^2$-Searcher. In contrast, the success rate of search and reasoning is enhanced significantly by SFT, while naive RAG suffers from irrelevant retrieved documents and inaccurate reasoning contexts. This shows that the fine-tuned base model can learn to generate high-quality formats, reasoning contents, and search queries instead of additional knowledge from trajectories executed by the teacher model. 

\section{Efficiency Comparison}
\label{efficiency}
In this section, we compare the efficiency of R$^2$-Searcher and AutoRefine using the average search steps, average token consumption per question, and average inference time per batch using the test set. Note that the token consumption includes the prompt tokens and the response tokens. 
The results for the two methods trained on 3B and 4B base models are shown in Table~\ref{tab:efficiency}.

\begin{table}
		\caption{\textbf{Efficiency comparison between R$^2$-Searcher and AutoRefine trained on 3B and 4B base models.}}
        \vspace{-2mm}
		\label{tab:efficiency}
        \renewcommand{\arraystretch}{0.97}
		\begin{tabular}{cccc}
			\toprule
			Methods & Avg. Steps & Avg. Tokens & Avg. Time\\
			\midrule
            AutoRefine-3B & 1.61 & 1230.9 & 88.40 ms\\
            R$^2$-Searcher-3B & 2.05 & 1703.5 & 106.9 ms \\
            \midrule
            AutoRefine-4B & 2.06 & 1649.8 & 129.1 ms\\
            R$^2$-Searcher-4B & 2.03 & 1691.2 & 133.1 ms \\
			\bottomrule
		\end{tabular}
        \vspace{-3mm}
\end{table}

The results show that on the 3B base model, AutoRefine trained on 3B model exhibit less search steps and token consumptions than R$^2$-Searcher, since AutoRefine-3B has a higher proportion of cases where it obtains the answer after just one search round compared to R$^2$-Searcher, yet its performance is lower than that of R$^2$-Searcher (shown in Table~\ref{tab:rq1_accuracy}), this indicates that AutoRefine suffers from under-search problem~\cite{zhang2026search}. Training on the 4B model, however, is notably different. AutoRefine-4B exhibits stronger search capability than its 3B version, with both methods achieving around 2.05 search steps. Additionally, R$^2$-Searcher-4B, with significantly higher search quality and reasoning quality, exhibits comparable token consumption and inference time to AutoRefine-4B.

\section{Data Efficiency in Tree-based Rollout}

Unlike chain-based RL methods that sample $M$ independent trajectories, tree-based rollout for combined actions, following the previous work Tree-GRPO~\cite{treegrpo}, constructs a tree structure where multiple rollouts share common prefixes. For a fixed token budget $B$, the expected cost for generating $n$ new branches from existing nodes is significantly lower than generating $n$ full trajectories from scratch. Assuming that the sampled nodes are uniformly distributed across the depths of the existing trajectories, the expected depth $\mathbb{E}[d]$ of a selected node is half of the maximum depth $D$:
\begin{equation}
    \mathbb{E}[d] = \frac{1}{D} \sum_{i=1}^{D} i = \frac{D+1}{2}.
\end{equation}
Since the cost of generating a trajectory grows linearly with its depth (i.e., cost $\propto$ steps), the remaining budget $B_{\text{rem}}$ required to complete the trajectory from depth $d$ is proportional to the remaining steps $(D - d)$. Therefore, the expected remaining budget is:
\begin{equation}
    \mathbb{E}[B_{\text{rem}}] = B \cdot \mathbb{E}\left[\frac{D - d}{D}\right] = B \left(1 - \frac{\mathbb{E}[d]}{D}\right) = \frac{B(D-1)}{2D}.
\end{equation}
This derivation confirms that each tree expansion incurs an average cost of $B(D-1)/2D$, rather than the full cost $B$ required for independent chain-based rollouts. Consequently, the total expected budget for tree-search sampling with $M$ initial trees and $n$ expansions per tree is given by $E[B_{\text{tree}}] = M \cdot B + \cdot n \cdot B(D-1)/2D$. 

This structure allows tree-based rollout to generate approximately $1.5\times$ more complete rollouts under the same computational budget compared to chain-based methods, thereby increasing the effective sample size for training without additional inference costs and enhancing the training efficiency.

\end{appendices}

\bibliographystyle{ACM-Reference-Format}
\bibliography{sample-base}

\end{document}